\documentclass[submission, Phys]{SciPost}

\graphicspath{{figs/}}
\usepackage{amsmath,amssymb}
\usepackage{physics}
\usepackage{bm}
\usepackage{color}
\usepackage{braket}
\usepackage{tikz}
\usetikzlibrary{arrows,positioning,calc,shapes.arrows,decorations.pathreplacing,shadows}
\usepackage{dsfont}
\usepackage{algorithm}
\usepackage{algpseudocode}
\usepackage{numprint}
\npthousandsep{\ensuremath\,\allowbreak}
\usepackage{hyperref}
\hypersetup{colorlinks=true,breaklinks,linkcolor=blue,urlcolor=blue,citecolor=blue}

\newcommand{\bx}{\ensuremath{\boldsymbol{x}}}
\newcommand{\by}{\ensuremath{\boldsymbol{y}}}
\newcommand{\bA}{\ensuremath{\boldsymbol{A}}}
\newcommand{\bu}{\ensuremath{\boldsymbol{u}}}
\newcommand{\bv}{\ensuremath{\boldsymbol{v}}}
\newcommand{\mrF}{\ensuremath{\mathrm{F}}}

\newcommand{\wmax}{\ensuremath{{\omega_\mathrm{max}}}}

\newcommand{\Gfour}{\ensuremath{{G}}}
\newcommand{\Gph}{\ensuremath{G}}
\newcommand{\calG}{\ensuremath{  {\mathcal{G}}  }}

\newcommand{\np}{\ensuremath{{n^\prime}}}

\newcommand{\SF}{\ensuremath{S^\mathrm{F}}}

\newcommand{\KF}{\ensuremath{K^\mathrm{F}}}

\newcommand{\UF}{\ensuremath{U^\mathrm{F}}}

\newcommand{\VF}{\ensuremath{V^\mathrm{F}}}

\newcommand{\enhUB}{\ensuremath{U^\mathrm{\overline{B}}}}

\newcommand{\mG}{\mathcal{G}}

\newcommand{\xzero}{\ensuremath{x^{(0)}}}
\newcommand{\xone}{\ensuremath{x^{(1)}}}
\newcommand{\xtwo}{\ensuremath{x^{(2)}}}
\newcommand{\xthree}{\ensuremath{x^{(3)}}}

\newcommand{\overlineB}{\ensuremath{{\overline{\mathrm{B}}}}}

\begin{document}

\begin{center}{\Large \textbf{
Sparse sampling and tensor network representation of two-particle Green's functions
}}\end{center}

\begin{center}
Hiroshi Shinaoka\textsuperscript{1*},
Dominique Geffroy\textsuperscript{2,3},
Markus Wallerberger\textsuperscript{4,3},
Junya Otsuki\textsuperscript{5},
Kazuyoshi Yoshimi\textsuperscript{6},
Emanuel Gull\textsuperscript{4},
Jan Kune\v{s}\textsuperscript{3,7}
\end{center}

\begin{center}
{\bf 1} Department of Physics, Saitama University, Saitama 338-8570, Japan
\\
{\bf 2} Department of Condensed Matter Physics, Faculty of
Science, Masaryk University, Kotl\'a\v{r}sk\'a 2, 611 37 Brno,
Czech Republic
\\
{\bf 3} Institute of Solid State Physics, TU Wien, 1040 Vienna, Austria
\\
{\bf 4} University of Michigan, Ann Arbor, Michigan 48109, USA
\\
{\bf 5} Research Institute for Interdisciplinary Science, Okayama University, Okayama 700-8530, Japan
\\
{\bf 6} Institute for Solid State Physics, University of Tokyo, Chiba 277-8581, Japan
\\
{\bf 7} Institute of Physics, Czech Academy of Sciences, Na Slovance 2, 182 21 Praha 8, Czech Republic
\\
* shinaoka@mail.saitama-u.ac.jp
\end{center}

\begin{center}
\today
\end{center}

\section*{Abstract}
{\bf
Many-body calculations at the two-particle level require a compact
representation of two-particle Green's functions. In this paper, we introduce
a sparse sampling scheme in the Matsubara frequency domain as well as
a tensor network representation for two-particle Green's functions. 
The sparse sampling is based on the intermediate representation
basis and allows  an accurate extraction of the generalized
susceptibility from a reduced set of Matsubara frequencies. The
tensor network representation
provides a system independent way
to compress the information carried by two-particle Green's functions.
We demonstrate efficiency of the present scheme for calculations of 
static and dynamic susceptibilities in single- and two-band Hubbard models 
in the framework of dynamical mean-field theory.}

\vspace{10pt}
\noindent\rule{\textwidth}{1pt}
\tableofcontents\thispagestyle{fancy}
\noindent\rule{\textwidth}{1pt}
\vspace{10pt}

\section{Introduction}
Two-particle (2P) Green's functions are building blocks of a variety of many-body theories~\cite{Mahan}.
They are a key element for calculation of
susceptibilities in the framework of dynamical mean-field theory
(DMFT)~\cite{Georges:1996un} as well as for diagrammatic calculations
including vertex corrections and diagrammatic extensions of
DMFT~\cite{Toschi:2007fq,Valli:2015di,Galler:2017bh,
	Rubtsov:2009bf,Hafermann:2009jy,Antipov:2014ij,Otsuki:2014ex,
	Kusunose:2006fp,Rohringer:2013in,Taranto:2014fw,Kitatani:2015da,Ayral:2015eha,Ayral:2016fo,Li:2015cwa}.
In equilibrium finite-temperature formalism, the 2P quantities depend on three Matsubara frequencies,
the low- and high-frequency parts of which must be treated accurately.
Storing 2P quantities alone is a challenge, more so when 
multiple orbitals, low symmetries or low temperatures are involved.

Several methods have been proposed to address the storage issue.
Conventional approaches are based on a separate treatment of the low- and
high-frequency
parts~\cite{Kunes:2011is, Rohringer:2012cc, Li:2016bn, Kaufmann:2017hw, Wentzell-arXiv}.
The frequency dependence is treated exactly in a small low-frequency box 
while in the outside region
an asymptotic form is used. 
This works efficiently at relatively high temperatures.
As the temperature is lowered the size of the low-frequency box grows, until it
becomes prohibitively large.

Recently, the intermediate representation (IR) basis was introduced as
a promising solution to the storage issue~\cite{Shinaoka:2018cg}. In
IR the size of the data grows only logarithmically
with the inverse temperature $\beta$ and the bandwidth. 
A fitting procedure allows IR expansion
of numerical data in the Matsubara frequency domain.
Nevertheless, two obstacles remain:
the computational
cost of the IR expansion and the size
of the IR tensor.

Regarding the first obstacle,
the input data for the fitting scheme is very large, having a
dense support in the Matsubara frequency domain. The fitting procedure
thus becomes prohibitively expensive at low $T$.
As for the second one, the IR represents a 2P quantity as a high-order
tensor involving spin and orbital dimensions, in addition to those for
the IR basis itself. Further compactification of the tensor is
required for solving realistic multi-orbital systems at low $T$.

In this paper, we address these two issues.
First, we introduce a sparse grid in the Matsubara frequency
domain, which contains the desired information about the 2P Green's functions
This extends the approach developed in
Ref.~\cite{Jia2019} for one-particle (1P) Green's functions.
We introduce an efficient tensor network representation of the IR tensor
and a fitting (regression) algorithm to determine it. 
Reduction of the data to be sampled thanks to the sparse grid
makes evaluation of the IR coefficients very efficient and solves the first issue.
The tensor regression provides a model-independent 
way to compress the IR tensor and tackles the second issue.

We demonstrate the performance of  
the present method in the context of DMFT.
First, we test the accuracy of sparse sampling and tensor
network representation by calculating the static susceptibility of 
the single-band Hubbard model on a square lattice.
Next, we show the efficiency of the present method for dynamical
susceptibility calculation for a two-band Hubbard model with low symmetry.

The paper is organized as follows.
In the next section, the IR
for 1P and 2P Green's functions is reviewed. Sparse sampling of
2P Green's functions is introduced in Sec.~\ref{sec:sparse_sampling}.
In Sec.~\ref{sec:dim_reduction},
the tensor network representation is presented. 
Its accuracy for computing of static
susceptibilities of the single-band Hubbard model is demonstrated in
Sec.~\ref{sec:single-band-hubbard-model}.
In Sec.~\ref{sec:two-band-hubbard-model}, we present
numerical results for dynamic susceptibility calculations, in the more
demanding context of an ordered phase of the two-band Hubbard
model. In Sec.~\ref{sec:summary}, we summarize and
conclude.

\section{Intermediate Representation (IR) for Green's functions}
Here we review the IR for 1P and 2P Green's functions introduced in~\cite{Shinaoka:2018cg} and \cite{Shinaoka:2017ix}.
The reader may refer to Section 7 of \cite{otsuki2019sparse} for a review.

\subsection{One-particle Green's function}
The IR for 1P Green's functions was introduced in
Ref.~\cite{Shinaoka:2017ix}. The spectral (Lehmann)
representation of the 1P Green’s function
$G
(\tau)$ in the imaginary-time domain reads
\begin{equation}
	G
	(\tau) = -\int^{\wmax}_{-\wmax}d\omega
	K^{\alpha}
	(\tau,\omega)\rho
	(\omega)\label{eq:spectral}
\end{equation}
where we assume $\hbar = 1$.
The superscript $\alpha$ specifies statistics: $\alpha=\mathrm{F}$ for fermion and $\alpha=\mathrm{B}$ for boson.
The spectrum $\rho(\omega)$ is assumed to be bounded within the interval $[-\wmax, \wmax]$.
The kernel $K^{\alpha}(\tau,\omega)$ reads
\begin{align}
	K^\alpha(\tau, \omega)
	&\equiv \omega^{\delta_{\alpha, \mathrm{B}}}
	\frac{e^{-\tau\omega}}{1 \pm e^{-\beta
			\omega}}
	\label{eq:K}
\end{align}
for $0 \leq \tau \leq \beta$ ($\beta$ is the inverse temperature).
Here, the $+$ and $-$ signs are used for fermions and bosons, respectively.
The extra $\omega$ factor for bosons in Eq. (\ref{eq:K})
is introduced in order to avoid a singularity of the kernel at $\omega=0$.

For fixed values of $\wmax$ and $\beta$,
the IR basis functions are defined through the singular value decomposition (SVD)
\begin{align}
	\label{eq:decomp-tau-omega}
	&K^\alpha(\tau,\omega) = \sum^{\infty}_{l=0}
	S^\alpha_l U^{\alpha}_l(\tau) V^{\alpha}_l(\omega),\\
	&\mathrm{with}~\int_0^\beta d \tau
	U_l^\alpha(\tau)U_{l^\prime}^\alpha(\tau) = \int_{-\wmax}^\wmax d
	\omega V_l^\alpha(\omega)V_{l^\prime}^\alpha(\omega) =
	\delta_{ll^\prime},\nonumber
\end{align}
where the singular values $S_l^\alpha$ ($>0$) decrease with increasing $l$ exponentially.

For fermions, the Green's function can be expanded as
\begin{align}
	G
	(\tau) &= \sum_{l=0}^\infty  
	\mG(l)U_l
	^\mrF(\tau)\label{eq:IR-decomp},\\
	\mG(l)
	&= - S_l^\mrF \rho_l
	,\label{eq:Gl_rhol}
\end{align}
where $\rho_l
\equiv \int_{-\wmax}^\wmax d\omega \rho
(\omega) V
^\mrF_l(\omega)$.
The exponential decay of $S_l
^\mrF$ ensures a fast decay of $G_l
$ if the spectrum is bounded in $[-\wmax, \wmax]$.
The accuracy of the expansion can be controlled by applying a cut-off for the singular values.
The Matsubara-frequency representation of the Green's function reads
\begin{align}
	G
	(i\omega_n)
	&= \int_0^\beta d \tau G
	(\tau) e^{i\omega_n \tau}
	= \sum_{l=0}^\infty 
	\mG(l)
	U
	^\mrF_l(i\omega_n)
	\label{eq:Gl-omega}\\
	&\mathrm{with}~U^\mrF_l(i\omega_n) \equiv \int_0^\beta d \tau
	U^\mrF_l(\tau) e^{i\omega_n \tau}.\nonumber
\end{align}

\subsection{General form of IR for Two-particle Green's functions}

The problem of the three- and four-point Green's functions was
considered in Ref.~\cite{Shinaoka:2018cg}. 
The four-point Green's function can be expressed in the Matsubara domain as
\begin{align}
	&G_{ijkl}(i\omega_n, i\omega_{n^\prime},i\nu_m) \nonumber\\
	& \equiv \beta^{-2}\int_0^\beta d\tau_1 d\tau_2 d\tau_3 d\tau_4
	e^{i(\omega_n \tau_{12} +  \omega_{n^\prime} \tau_{34} +  \nu_m \tau_{14})} \braket{T_\tau c^\dagger_i(\tau_1) c_j(\tau_2) c^\dagger_k(\tau_3) c_l(\tau_4)}\\
	& \equiv
	\sum_{r=1}^{16} \sum_{l_1 l_2 l_3} \calG_O(r, l_1, l_2, l_3)
	U^{\alpha}_{l_1}(i\omega) U^{\alpha'}_{l_2}(i\omega') U^{\alpha''}_{l_3}(i\omega''),
	\label{eq:ir-four-point2}
\end{align}
where $\omega$, $\omega'$, $\omega''$ stand for $r$-dependent combinations of 
$\omega_n$, $\omega_{n^\prime}$, and $\nu_m$ listed in Table~\ref{table:four-point}.
Similarly, the indices $\alpha$, $\alpha'$ and
$\alpha''$ take the value $F$ or $\bar{B}$ depending on $r$ as indicated in   Table~\ref{table:four-point}.
The indices $i$, $j$, $k$, $l$ denote the flavor (combined spin and
orbital), while $O$ is a composite index representing the quadruplet $(i,j,k,l)$.
Formally, the tensor
$\calG$ contains full information about
$G$ (meaning in particular, information about its values at all bosonic and fermionic frequencies for all flavors).

\begin{table}
	\centering
	\begin{tabular}{c|c|c|c}
		\hline
		$r$ &  ($i\omega$, $i\omega'$, $i\omega''$) & ($\alpha$, $\alpha'$, $\alpha''$) \\
		\hline
		\hline
1 & ($i\nu_m + i\omega_n$, $-i\omega_n$, $i\omega_{n^\prime}$) &  (F,F,F)  \\
2 & ($i\nu_m + i\omega_n$, $-i\omega_n$, $-i\nu_m - i\omega_{n^\prime}$) &  (F,F,F)  \\
3 & ($i\nu_m + i\omega_n$, $i\omega_{n^\prime}$, $-i\nu_m - i\omega_{n^\prime}$) &  (F,F,F)  \\
4 & ($-i\omega_n$, $i\omega_{n^\prime}$, $-i\nu_m - i\omega_{n^\prime}$) &  (F,$\overline{\mathrm{B}}$,F)  \\
5 & ($i\nu_m + i\omega_n$, $i\nu_m$, $i\nu_m + i\omega_{n^\prime}$) &  (F,$\overline{\mathrm{B}}$,F)  \\
6 & ($i\nu_m + i\omega_n$, $i\nu_m$, $-i\omega_{n^\prime}$) &  (F,$\overline{\mathrm{B}}$,F)  \\
7 & ($i\nu_m + i\omega_n$, $i\nu_m + i\omega_n + i\omega_{n^\prime}$, $i\nu_m + i\omega_{n^\prime}$) &  (F,$\overline{\mathrm{B}}$,F)  \\
8 & ($i\nu_m + i\omega_n$, $i\nu_m + i\omega_n + i\omega_{n^\prime}$, $i\omega_n$) &  (F,$\overline{\mathrm{B}}$,F)  \\
9 & ($i\nu_m + i\omega_n$, $i\omega_n - i\omega_{n^\prime}$, $-i\omega_{n^\prime}$) &  (F,$\overline{\mathrm{B}}$,F)  \\
10 & ($i\nu_m + i\omega_n$, $i\omega_n - i\omega_{n^\prime}$, $i\omega_n$) &  (F,$\overline{\mathrm{B}}$,F)  \\
11 & ($-i\omega_n$, $i\nu_m$, $i\nu_m + i\omega_{n^\prime}$) &  (F,$\overline{\mathrm{B}}$,F)  \\
12 & ($-i\omega_n$, $i\nu_m$, $-i\omega_{n^\prime}$) &  (F,$\overline{\mathrm{B}}$,F)  \\
13 & ($-i\omega_n$, $-i\omega_n + i\omega_{n^\prime}$, $i\nu_m + i\omega_{n^\prime}$) &  (F,$\overline{\mathrm{B}}$,F)  \\
14 & ($-i\omega_n$, $-i\nu_m - i\omega_n - i\omega_{n^\prime}$, $-i\omega_{n^\prime}$) &  (F,$\overline{\mathrm{B}}$,F)  \\
15 & ($i\omega_{n^\prime}$, $i\nu_m + i\omega_n + i\omega_{n^\prime}$, $i\nu_m + i\omega_{n^\prime}$) &  (F,$\overline{\mathrm{B}}$,F)  \\
16 & ($i\omega_{n^\prime}$, $-i\omega_n + i\omega_{n^\prime}$, $i\nu_m + i\omega_{n^\prime}$) &  (F,$\overline{\mathrm{B}}$,F)  \\
		\hline
	\end{tabular}
	\caption{16 different notations of Matsubara frequencies and
		statistics for the four-point Green's functions
                $\Gfour$. Adapted from Ref.~\cite{Shinaoka:2018cg}.
		}
	\label{table:four-point}
\end{table}

\subsection{Simplified form for fixed bosonic frequency}

\begin{figure}
  \begin{center}
    \resizebox{.45\columnwidth}{!}{%
      \begin{tikzpicture}[
        font=\sffamily,
        source/.style={draw,thick,rounded corners,fill=blue!20,inner sep=.3cm},
        ident/.style={draw,thick,inner sep=.3cm},
        process/.style={draw,thick,circle,fill=blue!20},
        every node/.style={align=center}]
        
        \node[draw,text width=2cm,minimum height=1cm,minimum
        width=4cm,thick,
        rounded corners,fill=blue!20,inner sep=.3cm, anchor=center] (full_correlator) {$\calG_O(r, l_1, l_2)$};
        \node[draw,text width=2cm,minimum height=1cm,minimum
        width=3cm,thick,
        rounded corners,fill=green!20,inner sep=.3cm,
        above left=1.3cm  and -1.5cm of full_correlator, anchor=center] (composed_object) {};
        
        \node[coordinate, above right=3.4cm and -1cm of composed_object] (fusion_center) {};
        \node[draw,minimum height=3cm,minimum width=6cm,thick,
        rounded corners,fill=green!20,inner sep=.3cm, anchor=center]
        at (fusion_center)  (detailed_object) {};
        
        \node[ident, path picture={%
          \draw (path picture bounding box.north west) --
          (path picture bounding box.south east);},
        below left=1cm and 2.2cm of fusion_center, anchor=center] (fusion) {};
        \node[ident, path picture={%
          \draw (path picture bounding box.north west) --
          (path picture bounding box.south east);},
        above=1cm of fusion_center, anchor=center] (fusion2) {};

        \draw[line width=0.6mm, dash dot] (composed_object.north west) --
        (detailed_object.south west);
        \draw[line width=0.6mm, dash dot] (composed_object.north east) --
        (detailed_object.south east);

        \node[process, below left=-0.1 cm and 0.5cm of fusion_center]
        (u1) {$U$};
        \node[process, below right=-0.1 cm and 0.5cm of fusion_center]
        (u2) {$U$};
        
        \draw[thick] (fusion.-90) -- node[pos=.6,left] {$r$} ++(0, -.8);
        \draw[thick] (fusion.0) -- node[midway,above] {$r$} (u1.-135);
        \draw[thick] (fusion.0) -- node[pos=0.75,below] {$r$} (u2.-135);
        \draw[thick] (u1.-90) -- node[pos=.8,left] {$l_1$} ++(0,-1.6);
        \draw[thick] (u2.-90) -- node[pos=.8,left] {$l_2$} ++(0,-1.6);
        
        \node[above=.07cm of fusion_center] (coordw) {$\mathcal{W}$};
        \draw[thick] (u1.55) -- (fusion2.-90);
        \draw[thick] (u2.125) -- (fusion2.-90);
        \draw[thick] (fusion2.90) -- node[pos=.8,left] {$\mathcal{W}$} ++(0,.8);
        
        \node[coordinate, above left=0cm and -2.5cm of full_correlator] (z) {};
        \node[coordinate, left=2cm of z] (x) {};
        \node[coordinate, left=1cm of z] (y) {};
        \node[coordinate, right=1cm of z] (zo) {};
        \draw[thick] (x) -- node[midway,left] {$r$} ++(0,.8);
        \draw[line width=0.3mm] (y) -- node[midway,left] {$l_1$} ++(0,.8);
        \draw[line width=0.3mm] (z) -- node[midway,left] {$l_2$} ++(0,.8);
        \draw[line width=0.3mm] (zo) -- node[pos=0.87,right] {$O$} ++(0,1.6);
        
        \draw[line width=0.3mm] (composed_object.45) -- node[midway,right] {$\mathcal{W}$} ++(0,.8);
        
      \end{tikzpicture}
    }%
  \end{center}
  
  \caption{
    Graphical representation for Eq.~(\ref{eq:ir-ph-tensor}).
    Each circle or rectangle with $N$ legs represents a $N$-way tensor.
    When two tensors share a leg, the summation over the corresponding index must be taken.
    A rectangle with a diagonal line represents an \textit{identity} tensor.
    The $N$-way \textit{identity} tensor $\boldsymbol{t}$ is
    defined as $t_{i_1 \cdots i_N}=1$ iff $i_1=\cdots=i_N$, and $t_{i_1
      \cdots i_N}=0$ otherwise.
  }
  \label{fig:tensor}
\end{figure}
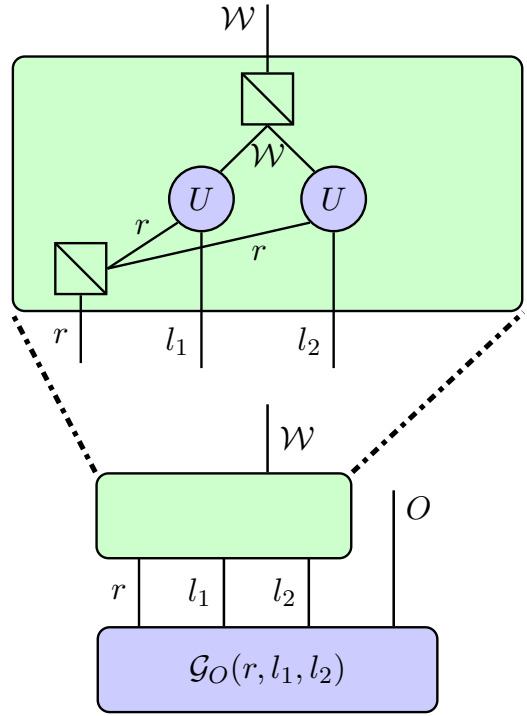

We now derive a variant of Eq.~(\ref{eq:ir-four-point2}), which holds
for a fixed bosonic frequency. The principle of the derivation is the
same as above. Nevertheless, if one is interested in only a
few bosonic frequencies, the reduced number of degrees of freedom
allows us to represent the 2P Green's function as a
tensor of lower rank.
In the conventional particle-hole notation,
the Green's function depends only on two fermionic frequencies for a fixed bosonic frequency.
It is
shown in Appendix~\ref{appendix:ph_view} that in this case, 
this frequency dependence can
be written as
\begin{align}
	&G_O(i\omega_n, i\omega_\np, i\nu_m)=\sum_{r=1}^{12}\sum_{l_1,l_2} \calG_O(r,l_1,l_2; i\nu_m)
	U_{l_1}^{(r)}(i\omega)U_{l_2}^{(r)}(i\omega'),
	\label{eq:ir-ph-tensor}
\end{align}
where $\omega_n$ and $\omega_\np$ are fermionic frequencies, and
$\nu_m$ is a bosonic frequency. 
The index $r$ relates to the 12 distinct representations generated
from the three terms in Eq.~(\ref{eq:ir-ph}).
The imaginary-time frequencies ($i\omega$,
$i\omega^\prime$) and the statistics of the basis functions depend on
$r$ as summarized in Table~\ref{table:four-point-ph}.

This expression
is formally similar to Eq.~(\ref{eq:ir-four-point2}),
being meant to store the full information for a single bosonic frequency.
The same sparse sampling strategy, which we shall 
introduce in Sec.~\ref{sec:sparse_sampling}, can therefore be
employed in both situations.

\begin{table}
	\centering
	\begin{tabular}{c|c|c}
		\hline
		$r$ & ($i\omega$, $i\omega$') & ($\alpha$, $\alpha'$) \\
		\hline
		\hline
		$1$ & ($i\omega_n$, $i\omega_{n^\prime}$) & (F, F) \\
$2$ & ($i\omega_n-i\omega_{n^\prime}$, $i\omega_{n^\prime}$) & (\overlineB, F) \\
$3$ & ($i\omega_{n^\prime}-i\omega_n$, $i\omega_n$) & (\overlineB, F) \\
\hline
$4$ & ($i\omega_n$, $i\omega_{n^\prime}+i\nu_m$) & (F, F) \\
$5$ & ($i\omega_n-i\omega_{n^\prime}$, $i\omega_{n^\prime}+i\nu_m$) & (\overlineB, F) \\
$6$ & ($i\omega_{n^\prime}-i\omega_n$, $i\omega_n+i\nu_m$) & (\overlineB, F) \\
\hline
$7$ & ($i\omega_n+i\nu_m$, $i\omega_{n^\prime}$) & (F, F) \\
$8$ & ($i\omega_n+i\omega_{n^\prime}+i\nu_m$, $i\omega_{n^\prime}$) & (\overlineB, F) \\
$9$ & ($i\omega_{n^\prime}+i\omega_n+i\nu_m$, $i\omega_n$) & (\overlineB, F) \\
\hline
$10$ & ($i\omega_n+i\nu_m$, $i\omega_{n^\prime}+i\nu_m$) & (F, F) \\
$11$ & ($i\omega_n+i\omega_{n^\prime}+i\nu_m$, $i\omega_{n^\prime}+i\nu_m$) & (\overlineB, F) \\
$12$ & ($i\omega_{n^\prime}+i\omega_n+i\nu_m$, $i\omega_n+i\nu_m$) & (\overlineB, F) \\
		\hline
	\end{tabular}
	\caption{12 different notations of Matsubara frequencies and
		statistics for four-point Green's functions in the
		particle-hole notation at a fixed bosonic frequency.
		}
	\label{table:four-point-ph}
\end{table}

\subsection{Graphical representation}
For the sake of clarity, we introduce a graphical representation for
the tensor operations involving 2P Green's functions.
As an example, we present in Fig.~\ref{fig:tensor} the diagram
corresponding to the right-hand side of Eq.~(\ref{eq:ir-ph-tensor}).
Each rectangle/circle represents a tensor whose indices are denoted by
legs (the nature of the shape does not matter).
The set of expansion coefficients $\calG_O(r,l_1,l_2; i\nu_m)$ appears
as a purple four-legged box, representing a rank four tensor.
The green rectangle represents the basis functions terms in
Eq.~(\ref{eq:ir-ph-tensor}).
A detailed view of their action is shown in the upper panel, where
$\mathcal{W}$ is introduced as a composite index of $(i\omega,
i\omega^\prime)$.

\section{Sparse sampling}\label{sec:sparse_sampling}
\begin{figure}
  \centering
  \includegraphics[width=0.7\linewidth]{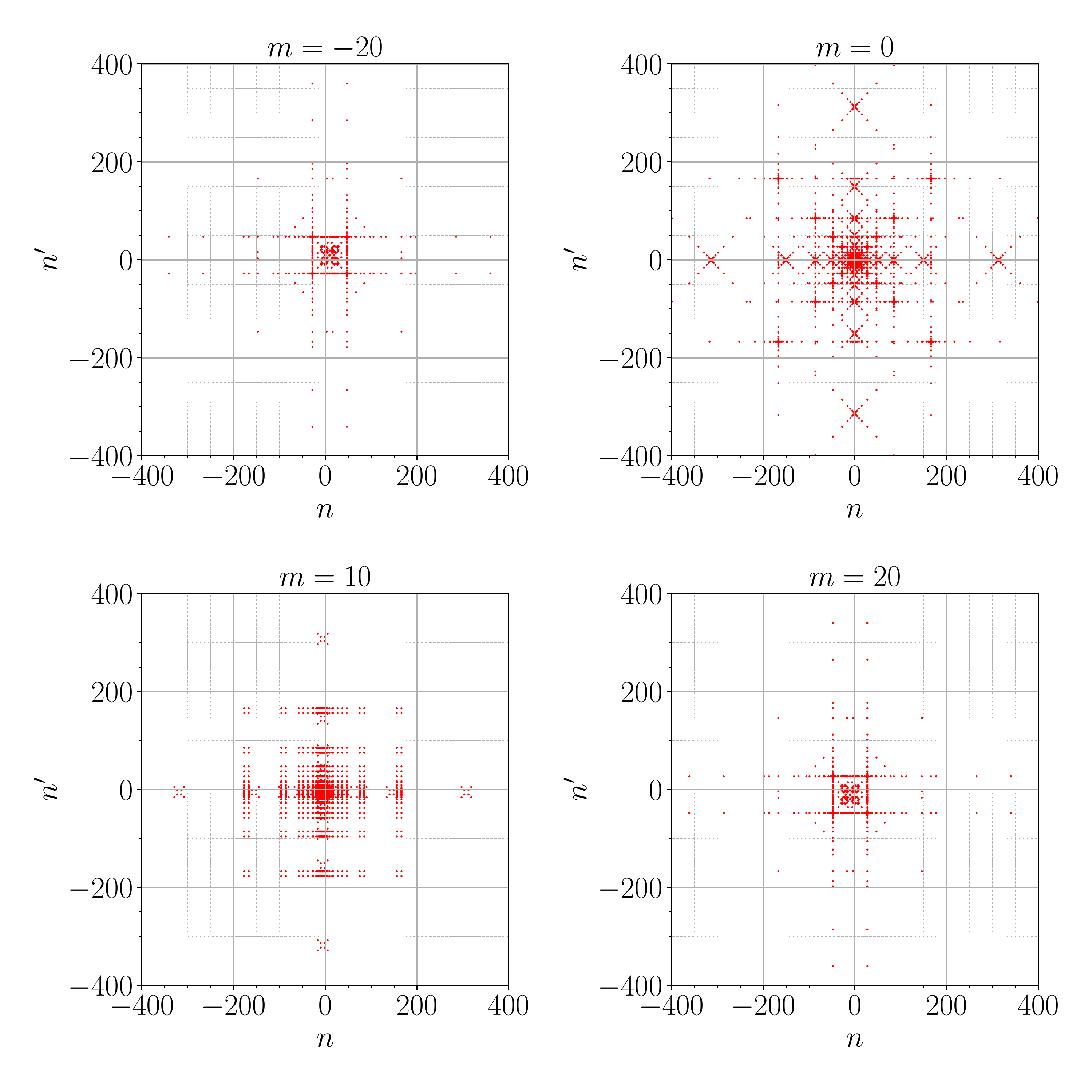}
  \caption{Sampling points generated in the three-frequency
    space for $\Lambda=10^4$ and $N_l=24$. We show cuts at
    $m=-20$, 0, 10, 20. We show only sampling points at low
    frequencies. The actual sampling points are distributed up
    to $|n|, |n^\prime| \lesssim 1800$ and $|m|\lesssim 2200$.}
  \label{fig:sampling_points_three_freqs}
\end{figure}

\begin{figure}
  \centering
  \includegraphics[width=0.7\linewidth]{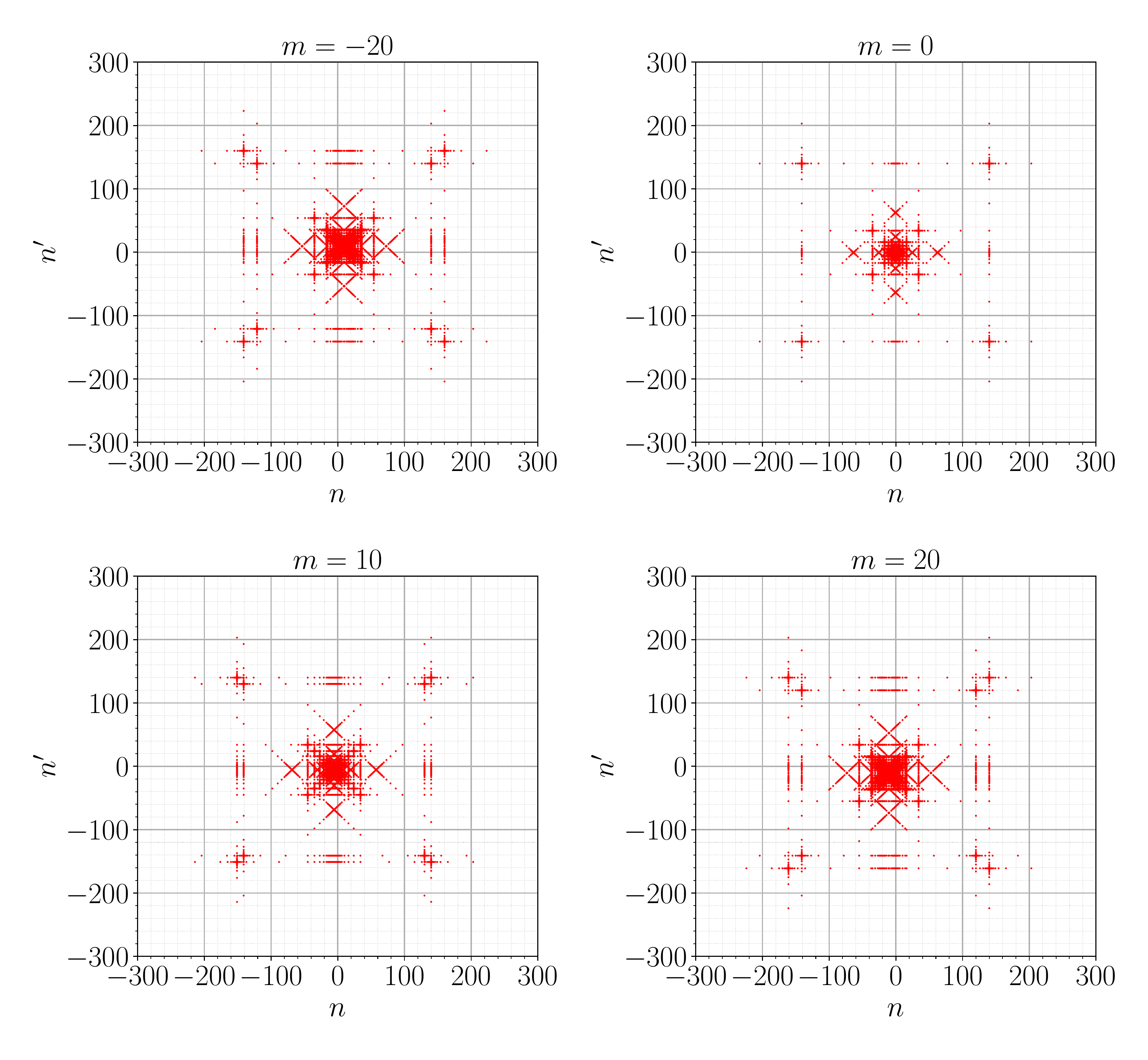}
  \caption{Sampling points generated in the particle-hole notation
    independently for $m=-20, 0, 10, 20$ ($\Lambda=100$ and $N_l=19$).}
  \label{fig:sampling_points_boson_freqs}
\end{figure}

The sparse sampling scheme was originally proposed for a
1P Green's function
in Ref.~\cite{Jia2019}.
For fermions, the expansion of
$G(i\omega_n)$ reads
\begin{align}
	G(i\omega_n) &= \sum_{l=0}^{N_l-1} \calG(l) U_l^\mrF(i\omega_n),
\end{align}
where the number of coefficients $N_l=l_\mathrm{max}+1$ determines the
accuracy of the expansion.
It was shown that 
the
full frequency dependence of a 1P Green's
function can be reconstructed from the values of the Green's function
on a carefully chosen sparse subset of sampling
points in the Matsubara domain.
$U_l^\mrF(i\omega_n)$ is real (odd $l$) or pure imaginary (even $l$), and oscillates around zero.
The procedure described in Ref.~\cite{Jia2019}  is based on picking
the positions of the extrema of
$|U_{l_\mathrm{max}}^\mrF(i\omega_n)|$.
The procedure generates $N_l$ (even $l$) or $N_l+1$ (odd $l$) sampling
points.
The same procedure generates  $N_l+1$ (even $l$) or $N_l$ (odd $l$) sampling points for bosons.

We extend this procedure in a straightforward manner for
the expansion of the 2P Green's function.
Each summand indexed by $r$
in Eq.~(\ref{eq:ir-four-point2})
(Eq.~(\ref{eq:ir-ph-tensor}) ) is handled in turn. For a fixed value
of $r$, the sets of sampling points relative to each factor in the
corresponding product of basis functions are built. Then, the triplets
(pairs) of direct products of such sets are determined, and make up
the $r^{\text{th}}$ set of points to be sampled, in $\mathds{Z}^3$
($\mathds{Z}^2$).

For simplicity of implementation, we use the same $N_l$ for
fermions and for bosons in the expansion. In general, for a given value
of $\Lambda \equiv \beta\wmax$, the singular values decay slower for fermions than for
bosons. In practice, we determine $N_l$ based on a given singular-value cutoff
for fermions, and use the same $N_l$ for bosons.

As an illustration, for Eq.~(\ref{eq:ir-four-point2}),
we obtain $N_l^3+O(1)$ sampling points.
The final set of sampling points we need
to consider is the union of the sets of sampling points obtained for
all $r$. The size of the union is less than the sum of the sizes of
the individual sets, thanks to the overlap between them (in
particular at low frequencies).
In Sec.~\ref{sec:two-band-hubbard-model}, we will use $\Lambda=10^4$ and $N_l=24$ (cutoff value $10^{-4}$ for singular values).
For this parameter set, the procedure generates \numprint{165912} sampling points for
Eq.~(\ref{eq:ir-four-point2}), which is slightly smaller than $16N_l^3 =
\numprint{221184}$ due to the overlap.
Figure~\ref{fig:sampling_points_three_freqs} shows the distribution of
these sampling points.
One can see that their distribution is more dense at low frequencies,
getting sparse at high frequencies.

Figure~\ref{fig:sampling_points_boson_freqs} shows the
distribution of the sampling points generated for
Eq.~(\ref{eq:ir-ph-tensor}) with $N_l=19$ (cutoff value of $10^{-5}$).
We obtain \numprint{2972},
1336, \numprint{2516}, \numprint{2972} sampling points for $m=-20, 0,
10, 20$, respectively.
We will use these parameters in Sec.~\ref{sec:single-band-hubbard-model}.

One technical caveat needs to be pointed out: the expansion
of the 2P Green's functions involves 
the so-called ``extended'' bosonic basis set\cite{Shinaoka:2018cg}.
A basis function $\enhUB_{l}$ from this set only exhibits 
$\text{max}(0, l - 2)$ sign changes, due to the extra basis functions
at $l=0, 1$. The procedure above would thus yield $l_\mathrm{max} - 1
= N_l - 2$ or $N_l-1$ sampling points for this basis set.
Therefore the actual process is 
slightly altered from the above description. The sampling
points relative to the extended bosonic basis are
generated from the extrema of $\enhUB_{l_\mathrm{max}+2}$
instead of $\enhUB_{l_\mathrm{max}}$. This ensures that the number of
unknown coefficients matches the number of sampling points.

In the
following sections, we will demonstrate that the sampling on the sparse grid is sufficient to
evaluate 
the 2P Green's function with the desired precision for any Matsubara frequency. 

\section{Tensor network representation\label{sec:dim_reduction}}

In this section, we introduce an efficient fitting algorithm based on
a tensor network representation for the IR tensor.
We refer the reader who is not familiar with tensor networks to Refs.~\cite{ck:2011gl,Bridgeman:2017ez,TensorsNet}.
In principle,
numerical data on the sampling points can be fitted using either
Eq.~(\ref{eq:ir-four-point2}) or Eq.~(\ref{eq:ir-ph-tensor}) by using
the least squares method. The computational load of this naive
approach scales as $O(N_l^6 N_\mathrm{orb}^4)$ or $O(N_l^9
N_\mathrm{orb}^4)$ for two- and three-frequency quantities,
respectively. Here $N_\mathrm{orb}$ is the number of orbitals, and $N_l$
grows logarithmically with respect to $\beta$. The
fitting rapidly becomes too costly at low temperature, especially for
three-frequency quantities.

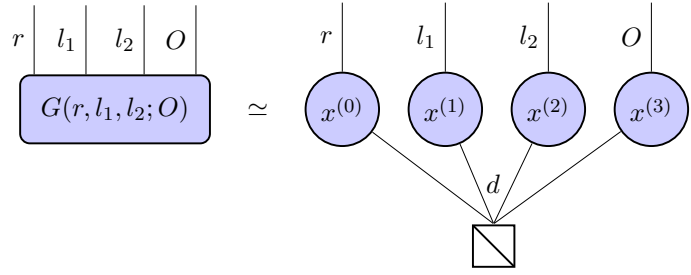
\begin{figure}
  \begin{center}
    \resizebox{0.6\columnwidth}{!}{%
      \begin{tikzpicture}[
        font=\sffamily,
        every matrix/.style={ampersand replacement=\&,column sep=.4cm,row sep=2cm},
        source/.style={draw,thick,rounded corners,fill=blue!20,inner sep=.3cm},
        ident/.style={draw,thick,inner sep=.3cm},
        process/.style={draw,thick,circle,fill=blue!20},
        every node/.style={align=center}]
        
        \matrix{
          \node[source] (full_correlator) {$\Gph(r, l_1, l_2; O)$};
          \& \node {$\simeq$};
          \& \node[process] (x0) {$x^{(0)}$};
          \& \node[process] (x1) {$x^{(1)}$};
          \& \node[process] (x2) {$x^{(2)}$};
          \& \node[process] (x3) {$x^{(3)}$};\\
        };
        \node[above=1.cm of full_correlator] (dummyc) {};
        \node[left=0.15cm of dummyc] (dummy) {};
        \node[left=0.90cm of dummyc] (dummyl) {};
        \node[right=0.15cm of dummyc] (dummyr) {};
        \node[right=0.90cm of dummyc] (dummyrr) {};
        \node[below=of dummy] (dummyb) {};
        \node[below=of dummyl] (dummybl) {};
        \node[below=of dummyr] (dummybr) {};
        \node[below=of dummyrr] (dummybrr) {};
        \draw (dummyb) -- node[midway,left] {$l_1$} (dummy);
        \draw (dummybl) -- node[midway,left] {$r$} (dummyl);
        \draw (dummybr) -- node[midway,left] {$l_2$} (dummyr);
        \draw (dummybrr) -- node[midway,left] {$O$} (dummyrr);
        
        \node[above=1.cm of x0] (dummyx0) {};
        \node[above=1.cm of x1] (dummyx1) {};
        \node[above=1.cm of x2] (dummyx2) {};
        \node[above=1.cm of x3] (dummyx3) {};
        \draw (x0) -- node[midway,left] {$r$} (dummyx0);
        \draw (x1) -- node[midway,left] {$l_1$} (dummyx1);
        \draw (x2) -- node[midway,left] {$l_2$} (dummyx2);
        \draw (x3) -- node[midway,left] {$O$} (dummyx3);
        
        \node[below=.7cm of x1] (dummyfusion) {};
        \node[ident, path picture={%
          \draw (path picture bounding box.north west) -- (path
          picture bounding box.south east);}, below right=0.15cm and .25cm of dummyfusion] (fusion) {};
        \node[circle,inner sep=0pt,minimum size=0.1pt, above=.0cm of fusion] (pointfusion) {};
        \draw (pointfusion) -- (x0);
        \draw (pointfusion) -- node[midway,right] {$d$} (x1);
        \draw (pointfusion) -- (x2);
        \draw (pointfusion) -- (x3);
      \end{tikzpicture}
    }%
  \end{center}
  \caption{
    Graphical representation of the tensor decomposition in Eq.~(\ref{eq:CP-ph}).
    Each circle or rectangle with $N$ legs represents a $N$-way tensor.
    When two tensors share a leg, the summation over the corresponding index must be taken.
    A rectangle with a diagonal line represents a \textit{identity tensor}.
  }
  \label{fig:cp}
\end{figure}

\subsection{Low-rank tensor decomposition}
We introduce the following low-rank decomposition of $\Gfour$:
\begin{align}
	\calG_O(r, l_1, l_2, l_3) \simeq \tilde{\calG}_O(r, l_1, l_2, l_3)\equiv\sum_{d=1}^D
	x^{(0)}_{dr} x^{(1)}_{dl_1} x^{(2)}_{dl_2}
	x^{(3)}_{dl_3} x^{(4)}_{dO},
	\label{eq:CP}
\end{align}
where $r$ runs over different representations and $O\equiv (i,j,k,l)$.
This type of tensor decomposition is known as a Canonical Polyadic
(CP) decomposition and is widely used in many
fields, e.g., for accelerating quantum chemistry calculations by factorizing Coulomb integrals~\cite{Hummel:2017fl,Motta:2019cd,Khoromskaia:2013dt}.
For the simplified form with fixed bosonic frequency in Eq.~(\ref{eq:ir-ph-tensor}), the low-rank decomposition reads
\begin{align}
	\calG_O(r, l_1, l_2;i\nu_m) \simeq
	\tilde{\calG}_O(r, l_1, l_2;i\nu_m)\equiv
	 \sum_{d=1}^D
	x^{(0)}_{dr}(\nu_m) x^{(1)}_{dl_1}(\nu_m) x^{(2)}_{dl_2}(\nu_m) x^{(3)}_{dO}(\nu_m),
	\label{eq:CP-ph}
\end{align}
which is  illustrated in Figure~\ref{fig:cp}.

Both expressions become exact for a large enough $D$.
The decomposition is beneficial for the fitting procedure if a
good approximation of the 
full tensor is obtained for a reasonably small value of $D$. We
discuss further in the text how this condition can be checked
numerically.
Note that the dependence on the orbital indices is not
decomposed and they still appear as the composite index $O$ in
Eqs.~(\ref{eq:CP}) and (\ref{eq:CP-ph}).
In the CP decomposition, we do not assume any orthogonality conditions for the decomposed
tensors, unlike the SVD of a matrix. Recently, some of the authors and
co-workers have proposed strong-coupling formulas for computing
momentum dependent susceptibilities in
DMFT~\cite{Otsuki:2019bz}. Their formulas can be regarded as a special
case of Eq.~(\ref{eq:CP-ph}) with $D=1$ and $x_{dr}=\delta_{r1}$.
Our preliminary results indicate that the further decomposition of the dependence on $O$ 
into individual spin/orbital indices
requires typically an even larger $D$, which is not beneficial.
\subsection{
Fitting algorithm}
We explain how to fit some existing data of the 2P Green's function on the sampling points in the
Matsubara domain using Eq.~(\ref{eq:CP}) or Eq.~(\ref{eq:CP-ph})
without explicitly constructing the big tensors on the left-hand
side.
The $\bx$ tensors in the
equation can be regarded as free fitting parameters. For instance, for
Eq.~(\ref{eq:CP}), we define a cost function for the fitting as
\begin{equation}
	f(\set{x^{(i)}})
	= 
	||G(W, O) -
	\tilde{G}(W, O)[x^{(i)}]
	||^2_2  
	+ \alpha \sum_{i=0}^4 ||x^{(i)}||_2^2,
	\label{eq:cost} 
\end{equation}
where $||\ldots||_2$ denotes the Frobenius norm
and $W$ runs over sampling points
in the Matsubara frequency domain.
$G(W, O)$ are the data of the Green's function on the sampling points which we fit, while $\tilde{G}(W, O)[x^{(i)}]$ denotes the data of the Green's function evaluated from $\set{x^{(i)}}$.
We introduce the small parameter $\alpha >0$ in
order to regularize the optimization problem. Without this
parameter, the problem would be ill-posed due to the overcompleteness
of the representation.
We have not observed any visible systematic errors in interpolated data for small values of $\alpha$, i.e., $\alpha\le 10^{-5}$.
We used $\alpha=10^{-8}$ and $10^{-5}$ in Sec.~\ref{sec:single-band-hubbard-model} and Sec.~\ref{sec:two-band-hubbard-model}, respectively.

The minimization of this cost
function is a non-convex optimization problem. 
We found that despite its non-convex nature, the cost function can be
minimized efficiently using standard methods starting from randomly initialized parameters.
In some cases, we observed the existence of multiple solutions,
being different only slightly in terms of the cost function.
This issue thus does not matter in practice.
We refer the interested reader to Appendix~\ref{appendix:ALS} for
more details on the optimization method.

\section{DMFT calculations for single-band Hubbard model on a square
  lattice}\label{sec:single-band-hubbard-model}
As a test bed for our method, we first consider a single-band
Hubbard model on a square lattice at half filling for 
$U=12t$ (1.5$\times$ the bandwidth $W=8t$), where the hopping $t=1$ 
sets the unit of energy.
The inverse temperature $\beta=2.5$ is slightly above the antiferromagnetic transition.
We use (approximate) Hubbard-I solver, which provides semi-analytic representation
of local susceptibilities and thus allows
precise analysis of our data compression approach.
We first compute and fit the local (impurity) generalized susceptibility $X^\mathrm{loc}$ by subtracting the relevant disconnected parts from 
the local 2P Green's function.
Interpolating the local generalized susceptibility in the Matsubara
frequency domain, we solve the Bethe-Salpeter equation (BSE) and
compute the static DMFT susceptibilities of the model.

\begin{figure}
  \centering
  \includegraphics[width=0.45\linewidth]{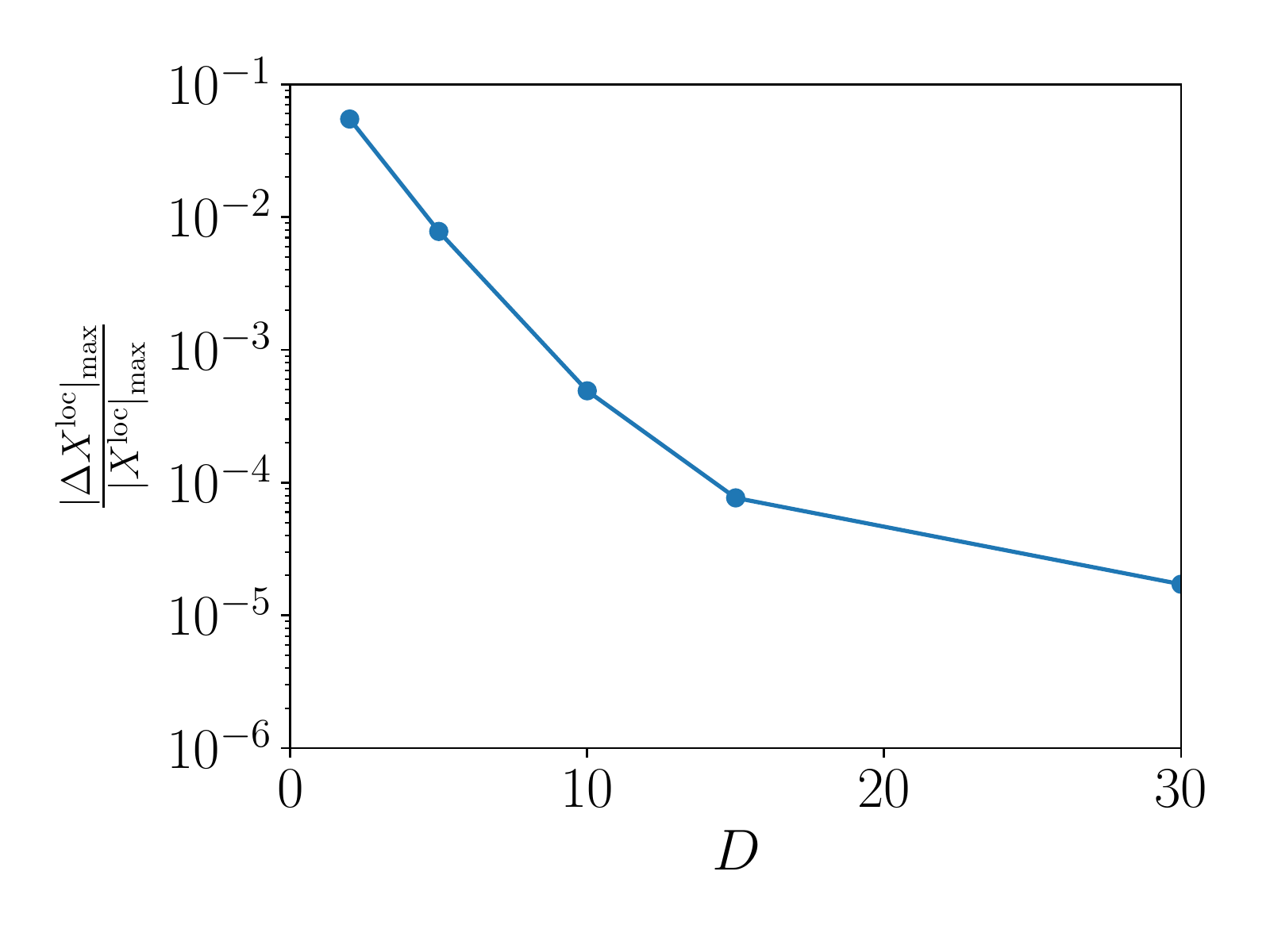}
  \caption{Fitting errors of $X^\mathrm{loc}$ for the single-band
    Hubbard model on the square lattice at $U=1.5W$ and $\beta=2.5$.
  }
  \label{fig:square_D_vs_max_error}
\end{figure}

\begin{figure}
  \centering
  \begin{minipage}[t]{0.4\linewidth}
    \begin{flushleft}(a) $O=(\uparrow,\downarrow,\uparrow ,\downarrow)$\end{flushleft}
    \includegraphics[width=\textwidth]{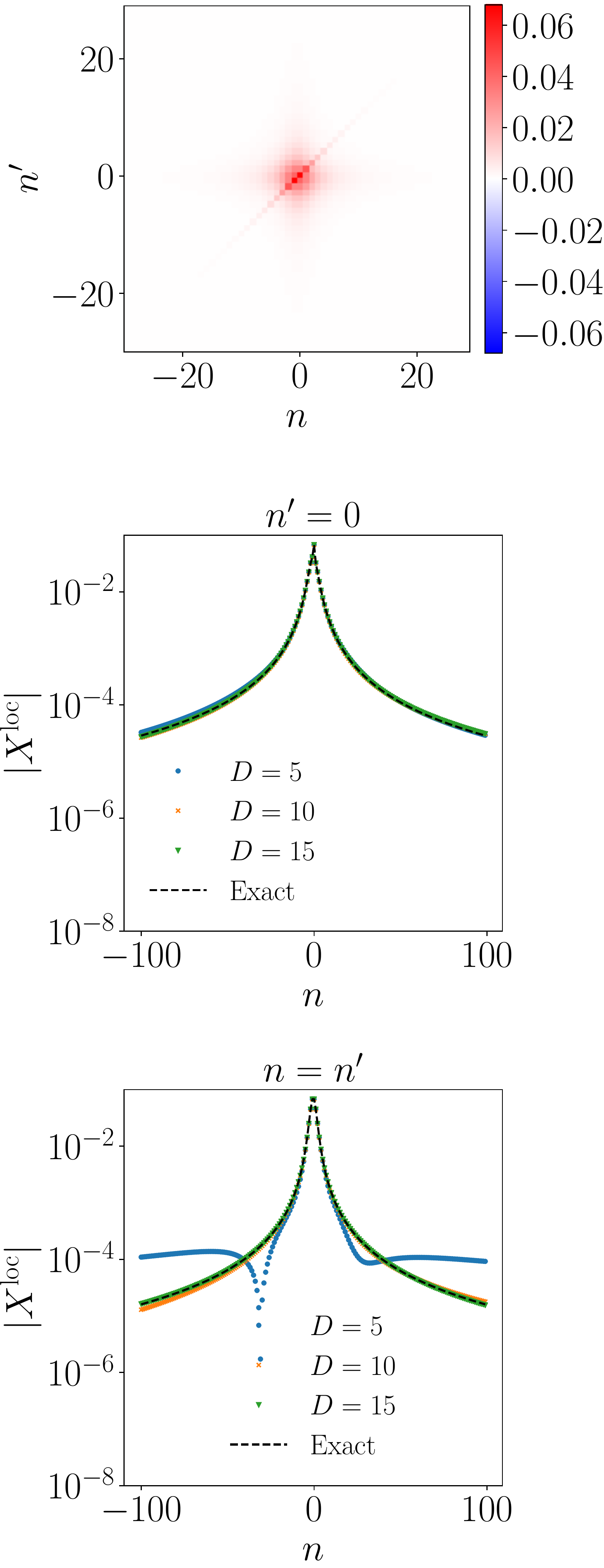}
  \end{minipage}
  \begin{minipage}[t]{0.4\linewidth}
    \begin{flushleft}(b) $O=(\uparrow,\uparrow,\uparrow ,\uparrow)$\end{flushleft}
    \includegraphics[width=\textwidth]{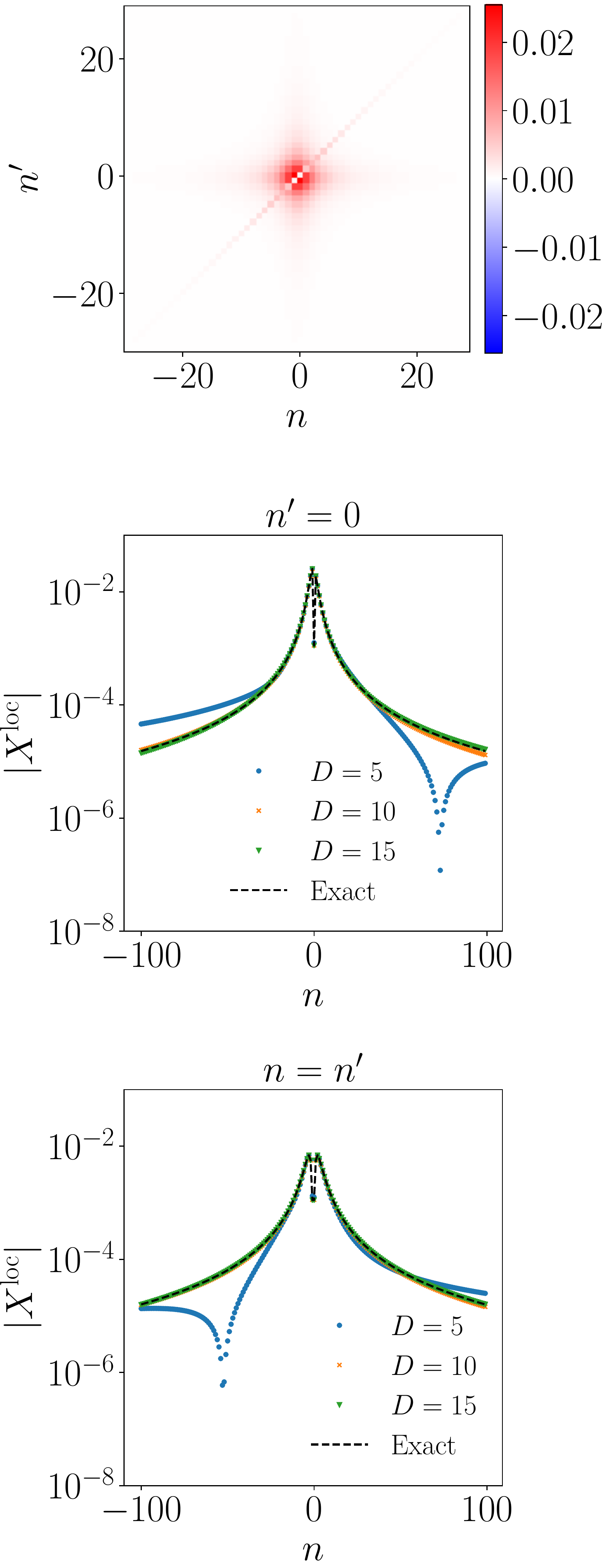}
  \end{minipage}
  \caption{
    Comparison of interpolated results and exact values
    of the generalized susceptibility $X^\mathrm{loc}$ for the single-band Hubbard model.
    The panels (a) and (b) show the data for
    $O=(\uparrow,\downarrow,\uparrow ,\downarrow)$ and
    $(\uparrow,\uparrow,\uparrow ,\uparrow)$, respectively.
    The 2D colormaps show the interpolated data for $D=15$.
  }
  \label{fig:comparison_Xloc}
\end{figure}

We compute  $X^\mathrm{loc}$ for
all spin-orbital components on \numprint{1336} sampling points
generated for $\Lambda=100$ and $N_l=19$ at zero bosonic frequency
$m=0$. The distribution of the sampling points is shown in the right
top panel of Fig.~\ref{fig:sampling_points_boson_freqs}.
Then, we fit the data using Eq.~(\ref{eq:CP-ph}) ($\Gph$ is
simply replaced by $X^\mathrm{loc}$).
Figure~\ref{fig:square_D_vs_max_error} shows how the fitting errors decay as $D$ is increased.
We found that the residual of the fit vanishes quickly with increasing $D$.
Figure~\ref{fig:comparison_Xloc} compares the exact and interpolated
values of the local susceptibility in Matsubara frequency space.
One can see that for $D=15$, the fit matches the exact values on the
sampling points and precisely interpolates the data. Increasing $D$
further does not improve the fit substantially, which may be due to
the truncation of the basis.

The inversion of the Bethe-Salpeter equations (for the determination of the
lattice susceptibilities) directly in the IR and tensor network
format is still an open question (see the discussion in
Sec.~\ref{sec:summary}). In this study, we execute the inversion
itself using the Matsubara representation, based on the interpolation of the
generalized susceptibility in a box of width $[-N_\omega,
N_\omega-1]$ for fermionic frequencies.
Corrections from higher frequencies outside the box are
treated using the procedure described in Appendix B of
Ref.~\cite{Otsuki:2019bz}. The corresponding physical quantities are
obtained by summation over the fermionic frequencies.

\begin{figure}
  \centering
  \includegraphics[width=0.45\linewidth]{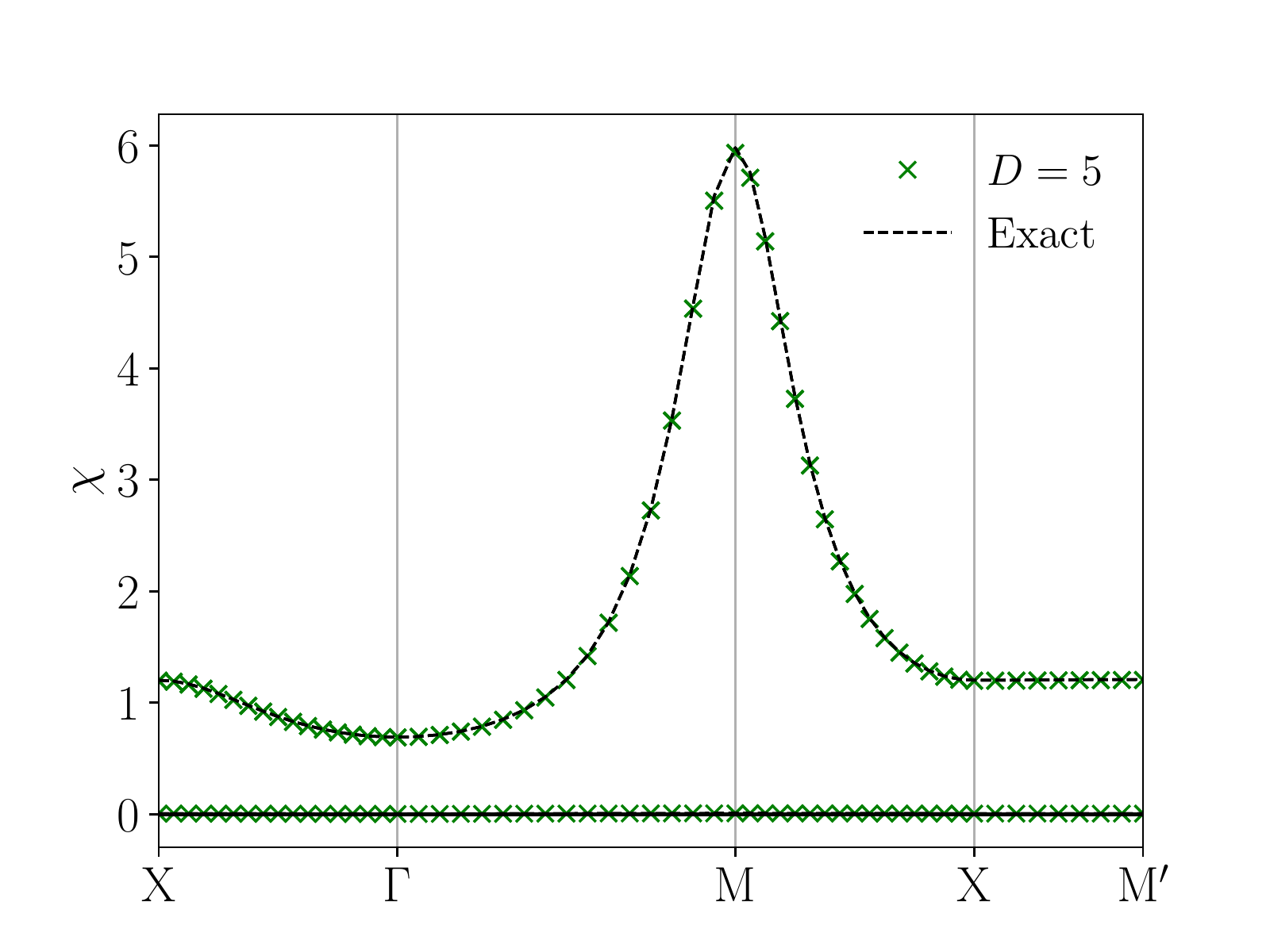}
  \caption{Physical lattice susceptibility $\chi(q)$ computed for the single-band
    Hubbard model on the square lattice at $U=1.5W$ and
    $\beta=2.5$. We used $D=5$. 
    Different branches correspond to different eigen modes of
    $\chi(\boldsymbol{q}, 0)$, i.e., the spin susceptibility and the
    charge susceptibility.
  }
  \label{fig:square_chiq_path}
\end{figure}

Figure~\ref{fig:square_chiq_path} shows the physical lattice
susceptibility using the fitted results for $D=5$ and $i\nu=0$.
For the calculations in this section, we took $N_\omega=100$. 
We found that the results for $D=5$ are already indistinguishable from
the exact ones at the scale of the figure. There is a pronounced peak
at $M=(\pi, \pi)$, corresponding to an antiferromagnetic spin
order. The other three eigenvalues, which correspond to charge
susceptibility, are not enhanced.

The storage of $X^\mathrm{loc}$ computed on the sampling points takes up
340 kB (including all spin sectors), while the compressed data for
$D=5$ only $5.2$ kB.

\section{DMFT calculations for two-band Hubbard model}\label{sec:two-band-hubbard-model}
Next, we apply the present IR approach to a state-of-art 
linear response DMFT calculation. To this end, we choose the two-band
Hubbard model in an intermediate coupling regime and low-spontaneously
broken-symmetry, a problem that some of us have studied
recently\cite{Geffroy:2019dh}.
The model Hamiltonian reads 
\begin{align}
	H=&\sum_{ij,\sigma} 
	 \left(t_{a}a_{i\sigma}^{\dagger}a_{j\sigma}^{\phantom\dagger}+
	 t_{b}b_{i\sigma}^{\dagger}b_{j\sigma}^{\phantom\dagger}\right)
	+\frac{\Delta}{2}\sum_{i,\sigma}(n^a_{i\sigma}-n^b_{i\sigma})
	\nonumber
	+ U \sum_{i,\alpha}n^\alpha_{i\uparrow}n^\alpha_{i\downarrow}+
	\sum_{i,\sigma\sigma'}(U'-J\delta_{\sigma\sigma'}) n^a_{i\sigma}n^b_{i\sigma'}, 
	\label{eq:2bhm}
\end{align}
where $a^{\dag}_{i\sigma}$ and $b^{\dag}_{i\sigma}$ are fermionic
operators that create electrons with the respective orbital flavors
and spin $\sigma$ at site $i$ of a square lattice. 
The first term describes the nearest neighbor hopping. The rest, expressed in terms
of local densities $n^{c}_{i,\sigma} \equiv
c^{\dag}_{i\sigma}c_{i\sigma}$, captures the crystal-field $\Delta$,
the Hubbard interaction $U$ and Hund's exchange $J$ in the Ising
approximation.
We use the same hopping parameters $t_{a}=0.4118$, $t_{b}=-0.1882$
as in~\cite{Geffroy:2019dh}, but choose weaker
interaction $U=2$ ($J=U/4$, $U'=U-2J$) and $\Delta=1.55$.
At the studied temperature $\beta=60$, an ordered phase called 
polar excitonic condensate~\cite{Kunes:2014ea} is realized. We follow
the same the algorithmic programme as in
Ref.~\cite{Geffroy:2019dh}, except for the representation of the
2P Green's function.
\begin{figure}
  \centering
  \begin{flushleft}\hspace{3em}(a) $O=(a\uparrow, a\uparrow, a\uparrow, a\uparrow), m = 0$ \end{flushleft}
  \includegraphics[width=0.45\linewidth]{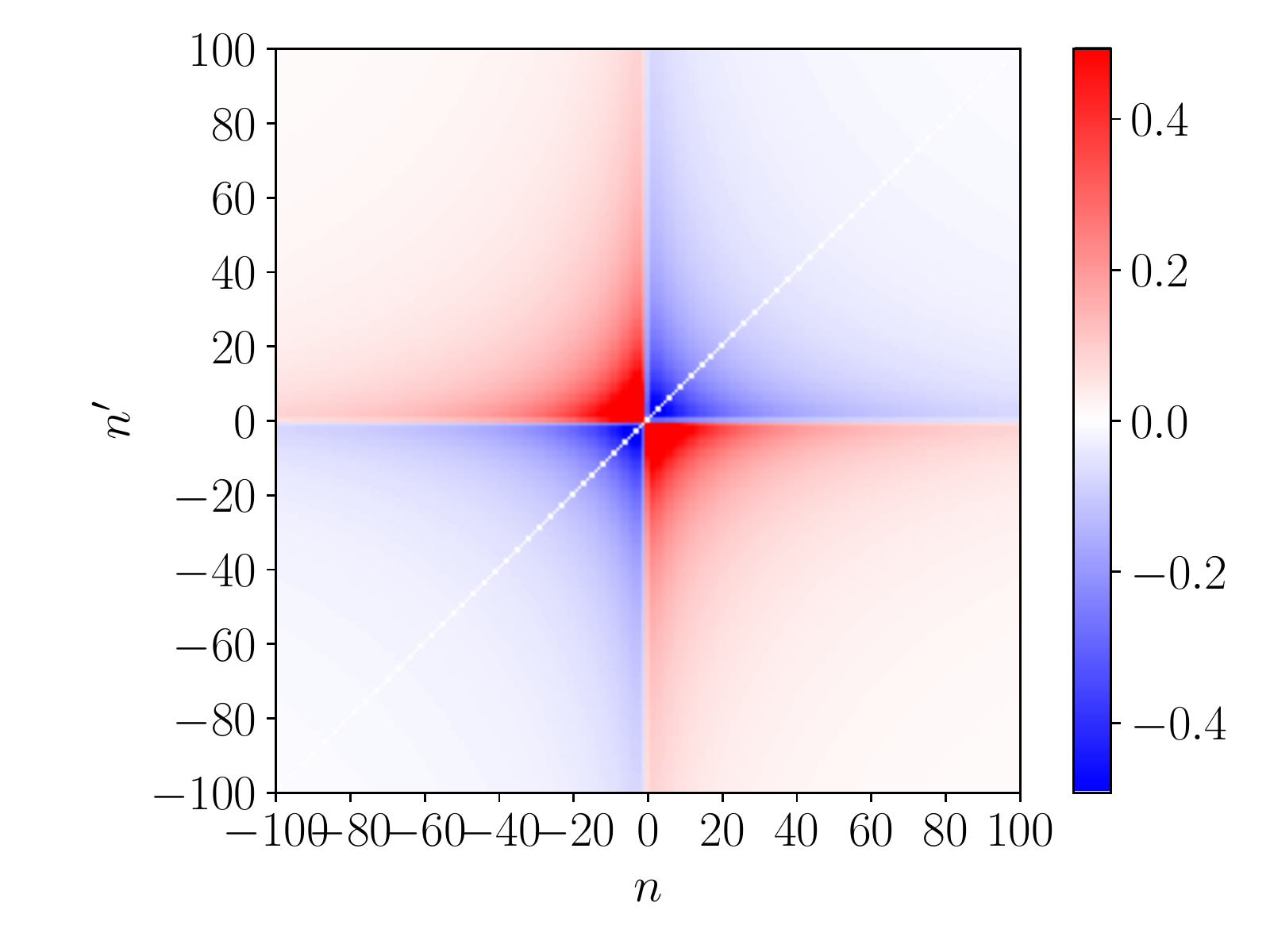}
  \includegraphics[width=0.45\linewidth]{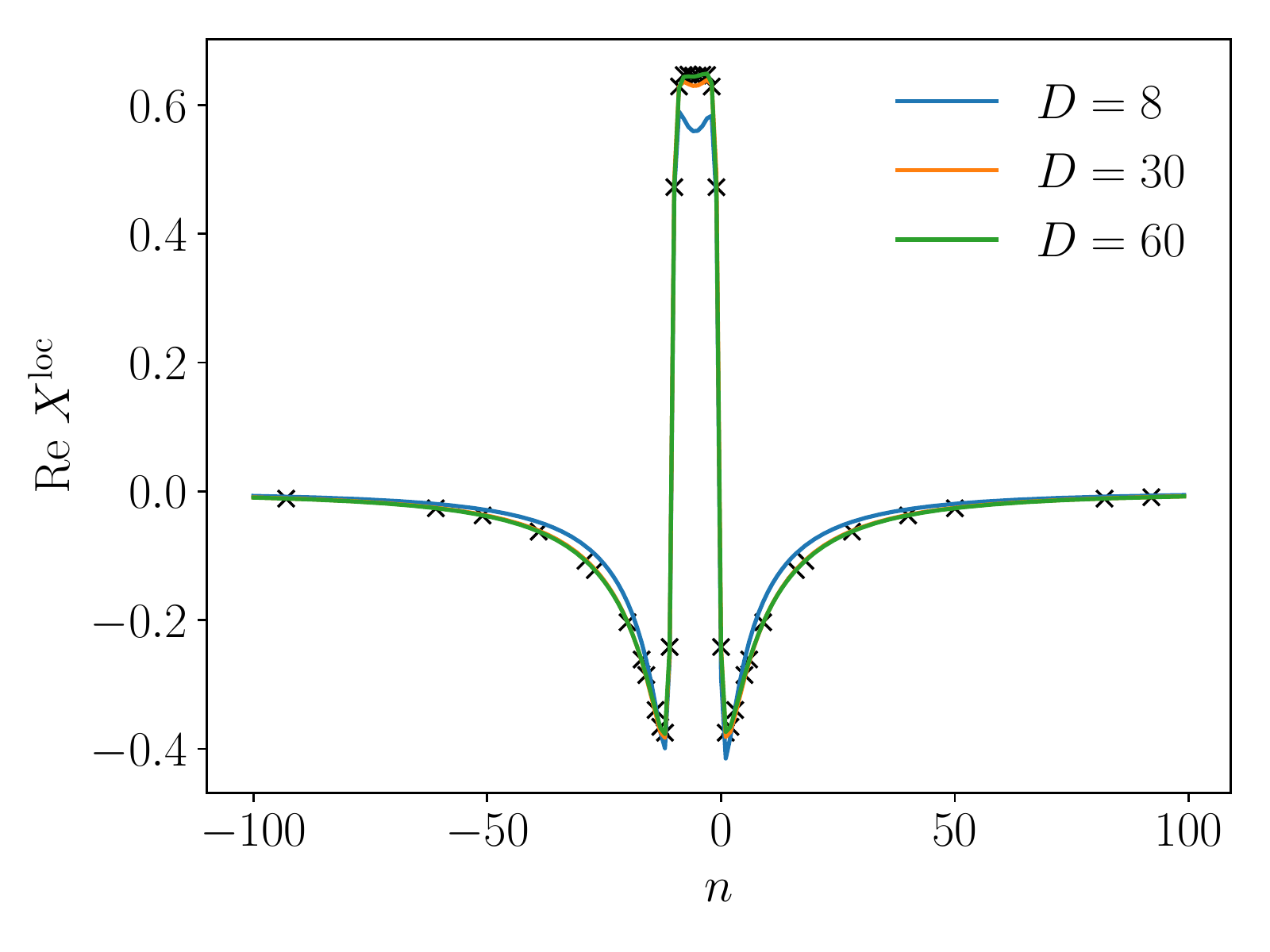}
  \begin{flushleft}\hspace{3em}(b) $O=(b\downarrow, b\downarrow, a\uparrow, a\uparrow)$ , m = 10\end{flushleft}
  \includegraphics[width=0.45\linewidth]{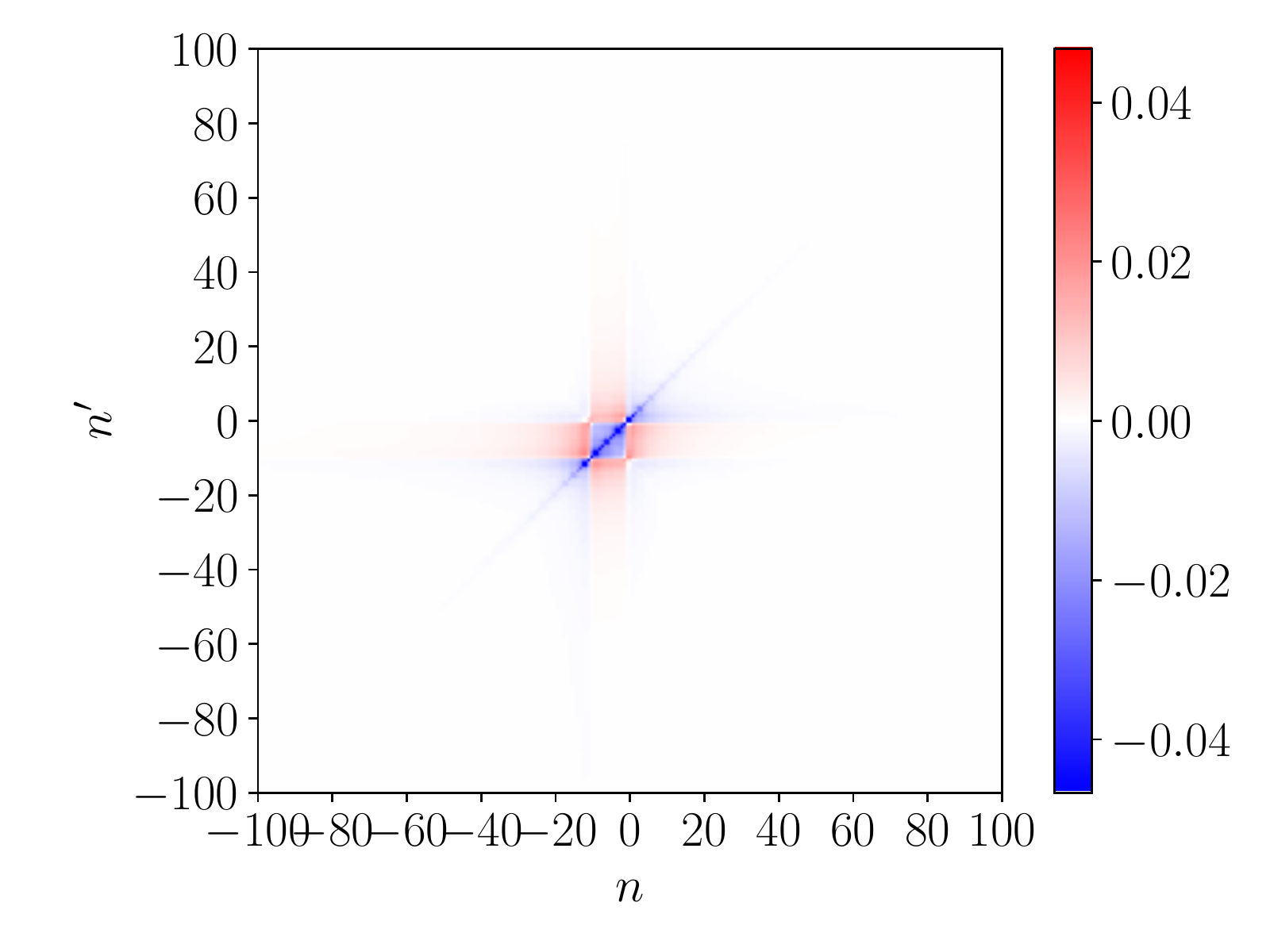}
  \includegraphics[width=0.45\linewidth]{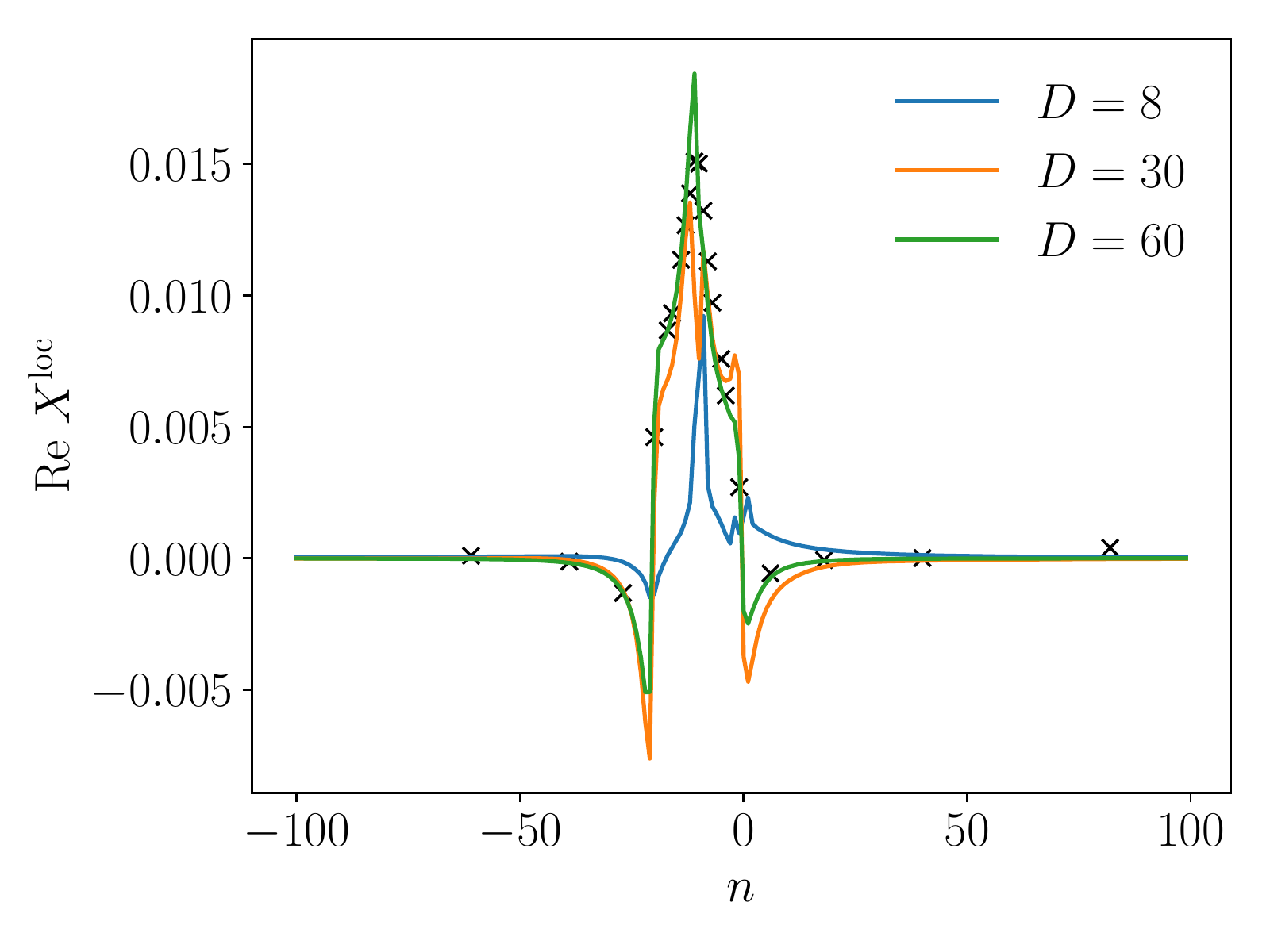}
  \caption{Local 2P Green's function $G^\mathrm{loc}$ interpolated at zero bosonic frequency
    $m=0$ [(a)] and a finite bosonic frequency $m=10$ [(b)] for the
    two-band Hubbard model.
    The 2D color maps show results of the local 2P Green's function for $D=60$.
    The right-hand halves of the panels (a) and (b) show the convergence of
    the interpolated values with respect to $D$ on the line of
    $n^\prime=n+10$.}
  \label{fig:comp_bose}
\end{figure}

The local 2P Green's function $G^\mathrm{loc}$ is sampled 
on a non-uniform grid in the Matsubara frequency domain, shown in
Fig.~\ref{fig:sampling_points_three_freqs} ($\Lambda=10^4$, $N_l=24$),
using a modified version of the ALPS/CT-HYB impurity
solver~\cite{Shinaoka:2017cpa,Shinaoka:2017ez} based on the
continuous-time hybridization expansion
algorithm~\cite{Werner:2006iz,Werner:2006ko}.
The  regression (\ref{eq:CP},\ref{eq:cost}) for $G^\mathrm{loc}$
provides us with the $\overline{x}^{(i)},\, i \in \{0, \ldots ,4
\}$ coefficients, Eq.~(\ref{eq:CP}), which in turn are used to
interpolate the local 2P Green's function at all Matsubara frequencies.
The generalized local susceptibility $X^\mathrm{loc}$ is then
evaluated by subtracting the disconnected part from the local
2P Green's function.

Figure~\ref{fig:comp_bose} illustrates the convergence of the
process with increasing value of $D$ in this situation. The solid
lines represent the real part of $G^\mathrm{loc}$ in the Matsubara representation, obtained from
Eq.~\ref{eq:ir-four-point2}. Both panels show the profile
for a fixed value of the bosonic frequency in the particle-hole
notation ($m=0$ and $m=10$), along a cut in the two-dimensional
fermionic frequency space, slightly away from the main diagonal
(shifted by ten Matsubara frequencies).
The crosses represent the
actual sparse frequencies, where the data was sampled. The
structure of the data at zero bosonic frequency is relatively simple,
so that the fit is excellent for $D=30$, but the tenth bosonic
frequency requires $D=60$.

\begin{figure}
  \centering
  \includegraphics[width=\linewidth]{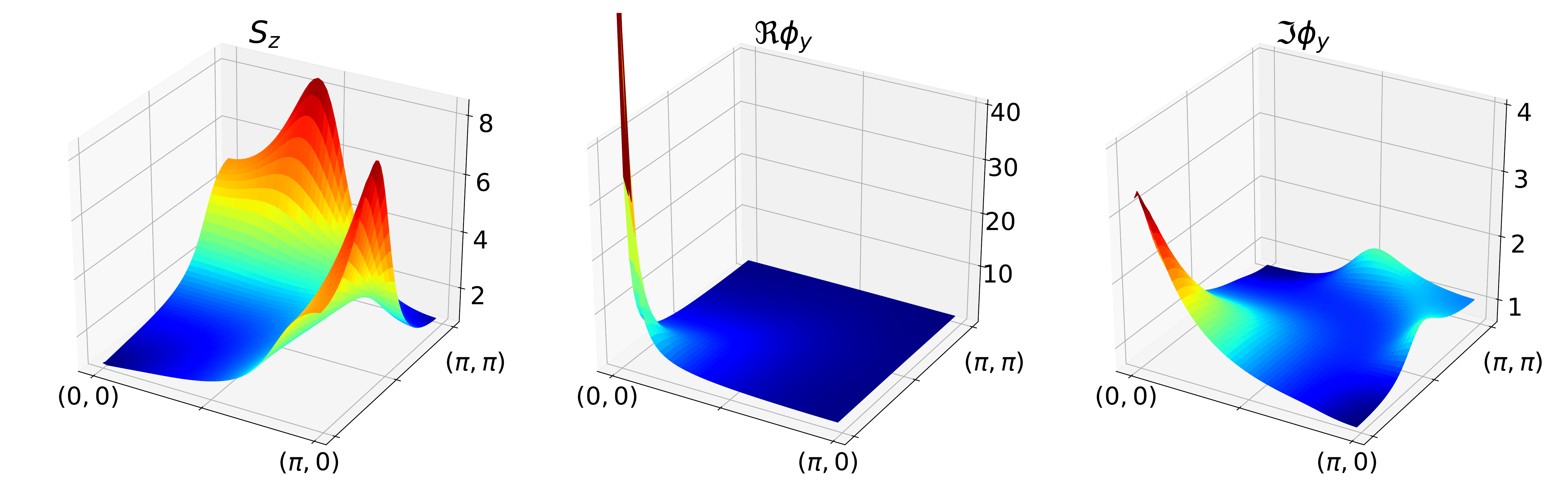}
  \caption{Selected components of the static susceptibility $\chi(\boldsymbol{q})$
  throughout
    the Brillouin zone, for the two-band Hubbard model described in
    the text, for $\beta=60$.}
  \label{fig:static_susc_MC}
\end{figure}

Using the above data we can solve the BSE to get the linear response
of our system. 
As an example we calculate the diagonal susceptibilities for local operators~\cite{Geffroy:2019dh}
\begin{gather*}
    S_z=\sum_{c=a,b}\left(n^c_{\uparrow}-n^c_{\downarrow}\right)\\
    \mathcal{R}\phi_\gamma=\sum_{\alpha\beta}\sigma^\gamma_{\alpha\beta}\left(a^{\dagger}_\alpha b^{\phantom\dagger}_\beta
    +b^{\dagger}_\alpha a^{\phantom\dagger}_\beta\right)\\
    \mathcal{I}\phi_\gamma=\sum_{\alpha\beta}i\sigma^\gamma_{\alpha\beta}\left(a^{\dagger}_\alpha b^{\phantom\dagger}_\beta
    -b^{\dagger}_\alpha a^{\phantom\dagger}_\beta\right),
\end{gather*}
which capture the low-energy dynamics of the polar condensate. We have chosen the ordered phase
such that only $\langle\mathcal{R}\phi_x\rangle$ is non-zero. In this set-up $\mathcal{R}\phi_y$ 
generates a spin rotation of the order parameter (Goldstone mode) and $\mathcal{I}\phi_y$ couples
to $S_z$ (only in the ordered phase).

\begin{figure}
	\centering
	\includegraphics[width=\linewidth]{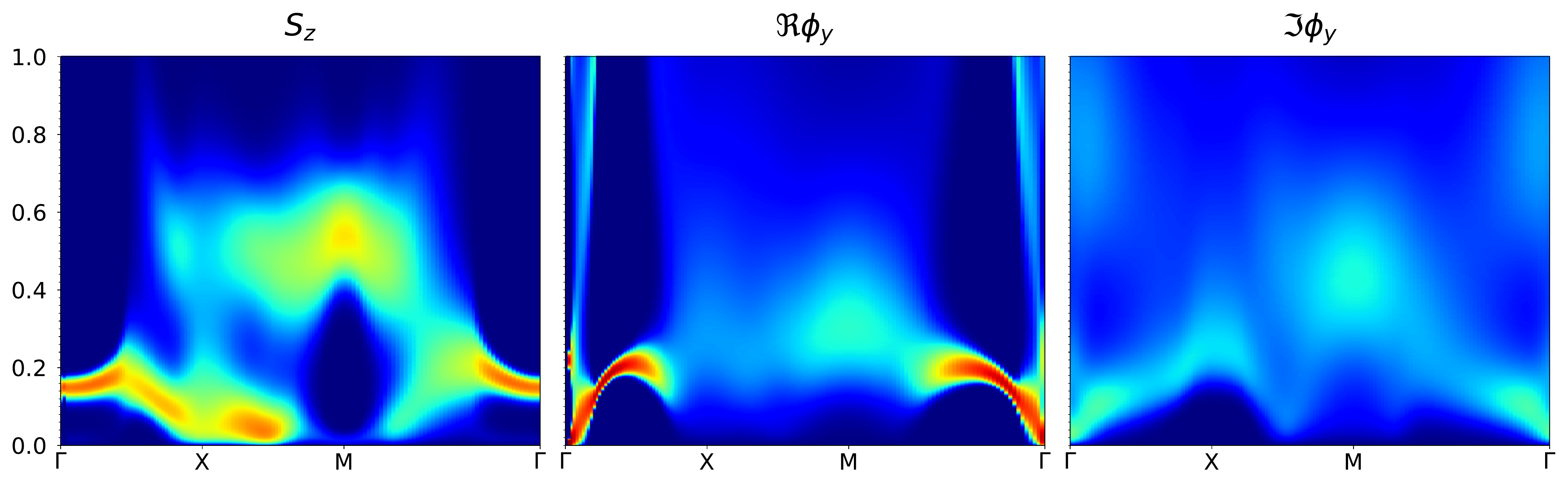}
	\caption{Selected modes of the dynamic susceptibility for the
          two-band Hubbard model described in the text, in the
          presence of the excitonic condensate, $\beta=60$.}
	\label{fig:dynamic_susc_MC}
\end{figure}

In Fig.~(\ref{fig:static_susc_MC}) we show the static susceptibilities on a fine grid in the 2D Brillouin zone.
Unlike in the strong coupling case of Ref.~\cite{Geffroy:2019dh}, the spin susceptibility is
dominated by a Fermi surface nesting present both in the ordered and
disordered (not shown here) phases, which gives rise to a peak at an
incommensurate vector on the $X$-$M$ zone boundary. The response
corresponding to the Goldstone
mode in the middle panel exhibits the expected divergence at the
ordering wave vector ($\mathbf{q}=0$).  The excitonic susceptibility
in the right panel exhibits, in addition to the main (finite) peak at
$\mathbf{q}=0$, additional peaks that coincide with the maxima of the
spin susceptibility. This reflects the coupling between $S_z$ and
$\mathcal{I}\phi_y$ induced by the symmetry breaking. While in the
strong coupling regime of Ref.~\cite{Geffroy:2019dh} $S_z$ was a slave
to the dynamics of $\mathcal{I}\phi_y$, here we can see that $S_z$
affects $\mathcal{I}\phi_y$.
In Fig.~(\ref{fig:dynamic_susc_MC}) we show the absorptive (imaginary)
parts of the dynamical susceptibilities obtained by the analytic
continuation described in Supplemental Material of
Ref.~\cite{Geffroy:2019dh}. It reveals the Goldstone nature of the
$\mathcal{R}\phi_y$ response and complex nature of the spin
response. Interestingly, the sharp response in the vicinity of
$\Gamma$ does not reflect formation of a bound state, but is a
consequence of parallel bands upon opening of the excitonic gap. The
low-energy hot spot on the $X$-$M$ linearand its counterpart in the
static susceptibility reflect the vicinity of an antiferromagnetic
phase ~\cite{Hoshino:2015gb}.

The presented susceptibilities  obtained
from the IR inputs are in excellent agreement with benchmark data
obtained using the Legendre representation used in Ref.~\cite{Geffroy:2019dh}. 
The current setup is more
flexible, insofar as it is not limited by the sampling window, neither
in the bosonic nor the fermionic Matsubara frequency domain.
It is also very compact.
The sparse grid solely saves computational time and memory footprints for QMC significantly.
The tensor regression further compresses the sparsely sampled data by
several orders of magnitude: The measured data is 700 MB large on the
sparse grid, while the tensor network representation takes up only 330
kB for $D=60$.

In general, if a too large $D$ is employed, one could overfit QMC noise,
giving a rise to oscillatory behavior between the sampling points in the interpolated
data.
A practical recipe for avoiding overfitting is to use the value
of $D$ that minimizes test errors rather than training errors.
In the present study, however, we did not observe overfitting behavior.
This may be because the fitting parameters is still much smaller than the fitted QMC data in size.
More detailed analysis of the stability of the fitting procedure is a topic of future studies.

\section{Summary}\label{sec:summary}
Based on the IR basis,
we have introduced a procedure for generating sparse grids in the Matsubara frequency
domain and a fitting algorithm based on a tensor network representation.
These two enable 
an efficient transformation of numerical data from Matsubara to IR domain.
The tensor network representation provides a model-independent way to
compress the IR expansion coefficients (IR tensor) by decoupling the
frequency and spins/orbital dependence.
Low-temperature calculations for multi-orbital systems benefit from this compression.

We have demonstrated the efficiency and accuracy of the present method
in DMFT calculations: static susceptibility calculations for
single-band Hubbard model and dynamic susceptibility calculations for
two-band Hubbard model with low symmetry.
We have shown that accurate susceptibilities can be obtained already
with low-rank approximation of the IR tensor.

The sparse sampling and the tensor network decomposition are independent procedures that are controlled separately.
The size of sparse sampling grid, and thus its computational cost, depends only on temperature, the energy window and the desired accuracy. 
The ``compression rate'' and the accuracy of tensor network representation are controlled by the rank of decomposition $D$.
In the present work, we have demonstrated that the local 2P Green's function can be compressed from 700 MB to 330 kB for the two-band Hubbard model.
The concept of tensor network representation is flexible and further 
compression may be possible for different tensor network topology.
The choice of ideal tensor network topology requires an extensive
experience with the performance of the method for various models and
is beyond the scope of the present work.

Potential applications of the present scheme include DFT+DMFT
calculations for realistic multi-orbital models and diagrammatic
extensions of DMFT.
It is highly desirable to develop efficient methods for solving
equations at the 2P level such as Bethe-Salpeter and parquet equations
directly in the tensor network format with the sparse sampling.
This requires efficient evaluation of contractions of 2P quantities,
e.g., a vertex function and a generalized susceptibility.
Potentially useful techniques for manipulating matrix product states
and tensor networks have already been developed in other fields of
condensed matter theory~\cite{ck:2011gl,Bridgeman:2017ez,Liao:2019en}.

\section*{Acknowledgments}
HS thanks Lei Wang for stimulating discussions on tensor networks and machine learning.
Part of the calculations were run on the facilities of the
Supercomputer Center at the Institute for Solid State Physics,
University of Tokyo, using codes based on
ALPSCore~\cite{Gaenko2018,Wallerberger:2018uf}. We used the
\texttt{irbasis} library~\cite{irbasis2019} for computing IR basis
functions. We used DCore~\cite{DCore} based on TRIQS~\cite{Parcollet:2015gq} and TRIQS/DFTTools~\cite{Aichhorn:2016cd} for DMFT calculations of the single-band Hubbard model.

\paragraph{Funding information}
H.S, J.O and K.Y were supported by JSPS KAKENHI Grant No. 18H01158.
H.S. was supported by JSPS KAKENHI Grant No. 16K17735.
J.O. was supported by JSPS KAKENHI Grant No. 18H04301 (J-Physics).
K.Y. was supported by Building of Consortia for the Development of
Human Resources in Science and Technology, MEXT, Japan.
D.G and J.K were supported by the ERC Grant Agreements No. 646807 under
EU Horizon 2020.
D.G was supported by the Czech Science Foundation (GA\v{C}R) under
Project No.~GA19-16937S.
This work was supported by The Ministry of
Education, Youth and Sports from the Large Infrastructures for
Research, Experimental Development and Innovations project
"IT4Innovations National Supercomputing Center – LM2015070".
This work was supported by The Czech Ministry of Education, Youth and
Sports from the Large Infrastructures for Research, Experimental
Development and Innovations projects ``IT4Innovations National
Supercomputing Center – LM2015070''. MW and EG were supported by the Simons Foundation via the Simons Collaboration on the many-electron problem.
Access to computing and storage facilities owned by parties and projects contributing to the National Grid Infrastructure MetaCentrum provided under the programme ``Projects of Large Research, Development, and Innovations Infrastructures'' (CESNET LM2015042), and by the Austrian Federal Ministry of Science, Research  and  Economy through the Vienna Scientific Cluster (VSC) Research Center, is greatly appreciated.

\begin{appendix}
\section{Intermediate representation at fixed bosonic frequency\label{appendix:ph_view}}
The frequency dependence of $\Gfour(i\omega_n, i\omega_{n^\prime}, i \nu_m)$ can be
decomposed into 16 distinct components shown in Table.~\ref{table:four-point}.
For instance, the first component ($r=1$) depends on the three frequencies through $i\nu_m + i\omega_n$, $-i\omega_n$, $i\omega_{n^\prime}$,
which defines the structure of discontinuity planes in the imaginary-time domain.
To be more specific, as discussed in Ref.~\cite{Shinaoka:2018cg}, the first component can be discontinuous at three equal-time planes: $\tau_1 = \tau_4$, $\tau_2 = \tau_4$ and  $\tau_3 = \tau_4$ (mod $\beta$).
Such a function with this discontinuity structure may be well approximated by
\begin{align}
\frac{\rho^{(1)}(\epsilon_1, \epsilon_2, \epsilon_3)}{(i\nu_m + i\omega_n - \epsilon_1) (-i\omega_n - \epsilon_2) (i\omega_{n^\prime} - \epsilon_3)},
\end{align}
where $\rho^{(1)}(\epsilon_1, \epsilon_2, \epsilon_3)$ is an \textit{auxiliary} spectrum bounded in $[-\wmax, \wmax]$.
Applying the same procedure to all the 16 components,
we obtain the assumption that 
\begin{align}
\Gfour(i\omega_n, i\omega_{n^\prime}, i \nu_m) & \simeq \sum_{r=1}^{16} \int_{-\wmax}^\wmax d\epsilon_1 d\epsilon_2 d \epsilon_3
\frac{\rho^{(r)}(\epsilon_1, \epsilon_2, \epsilon_3)}
{(i\omega-\epsilon_1)(i\omega'-\epsilon_2)(i\omega''-\epsilon_3)}.
\label{eq:ir-four-spectrum}
\end{align}

For instance, the term for $r=1$ can be decomposed as
\begin{align}
&\frac{1}{
	(i(\nu_m + \omega_n)-\epsilon_1)
	(-i\omega_n-\epsilon_2)
	(i\omega_{n^\prime}-\epsilon_3)
}\nonumber\\
&=
\frac{1}{i\nu_m-\epsilon_1-\epsilon_2}
\left(
\frac{1}{i(\nu_m+\omega_n)-\epsilon_1}
+\frac{1}{-i\omega_n-\epsilon_2}
\right)
\frac{1}{(i\omega_{n^\prime}-\epsilon_3)}\nonumber\\
&=\frac{1}{i\nu_m-\epsilon_1-\epsilon_2}
\left\{
\KF(i\omega_n + i\nu_m, \epsilon_1)-
\KF(i\omega_n + \epsilon_2)
\right\}
\KF(i\omega_{n^\prime}, \epsilon_3).
\end{align}
One can see the first and second terms in the last line are the products of two fermionic kernels.
The first term can be represented compactly as
\begin{align}
&\sum_{l_1,l_2}\left(\frac{\SF_{l_1}\SF_{l_2}
	\VF_{l_1}(\epsilon_1)\VF_{l_2}(\epsilon_3)}{i\nu_m-\epsilon_1-\epsilon_2}\right)
\UF_{l_1}(i\omega_n+i\nu_m)
\UF_{l_2}(i\omega_{n^\prime}),
\end{align}
where the coefficient in the parenthesis decays as fast as the
singular values with respect to $l_1$ and $l_2$.

Applying the same procedure to all the terms in Eq.~(\ref{eq:ir-four-spectrum}),
one obtains a compact overcomplete representation 
\begin{align}
\Gph(i\omega_n, i\omega_\np, i\omega_m) & =  \sum_{s,s^\prime=0,1}\sum_{l_1,l_2=0}^\infty \Big\{\calG^{(1)}_{s s^\prime l_1l_2}
\UF_{s l_1}(i\omega_n)\UF_{s^\prime l_2}(i\omega_\np) \nonumber\\
& \quad+  \calG_{s s^\prime l_1 l_2}^{(2)} \enhUB_{l_1 s}(i\omega_n +
(-1)^{s+1} i \omega_\np)\UF_{l_2 s^\prime}(i \omega_\np)
\nonumber\\
& \quad+
\calG_{s s^\prime l_1 l_2}^{(3)} \enhUB_{l_1 s}(i\omega_\np +
(-1)^{s+1} i\omega_n)
\UF_{l_2 s^\prime}(i\omega_n) \Big\}.
\label{eq:ir-ph}
\end{align}
Here we defined
\begin{align}
U_{s l}^\alpha(i\omega_n) &\equiv U_{ l}^\alpha(i\omega_n + s i\omega_m)
\end{align}
for $s=0, 1$.
The sum can be restricted to $s,s^\prime=0$ in the case of zero
bosonic frequency, i.e., $\nu=0$. In the present study, for simplicity
of the implementation, we also keep $s,s^\prime=1$ in the sum also for
$\nu=0$, which does not harm.

We can recast Eq.~(\ref{eq:ir-ph}) into a form more analogous to Eq.~(\ref{eq:ir-four-point2}) as
\begin{align}
G(i\omega_n, i\omega_\np, i\omega_m) &=\sum_{r=1}^{12}\sum_{l_1,l_2} \calG(r,l_1,l_2;i\nu)
U_{l_1}^{(r)}(i\omega)U_{l_2}^{(r)}(i\omega').
\end{align}


\section{Optimization algorithm for tensor regression\label{appendix:ALS}}
We minimize the cost function in Eq.~(\ref{eq:cost}) by means of a
accelerated alternating least squares (ALS) method. The essential idea
of ALS is to optimize each tensor in $\{ x^{(0)}, x^{(1)},\cdots \}$
at one time. The optimization of a single tensor reduces to a convex
optimization problem. In ALS, we sweep through all the tensors until
the value of the cost function is converged. In addition, we
introduce recently proposed acceleration techniques to improve the
convergence of ALS. In the following, we detail the procedure of ALS
and the acceleration techniques.

\subsection{Alternating least squares}
We explain how to optimize the tensors in Eq.~(\ref{eq:cost}) by alternating least squares.
\subsubsection{Optimization of $\xzero$}
The minimization of Eq.~(\ref{eq:cost}) with respect to $\xzero$ can be recast into
\begin{align}
\underset{\xzero}{\mathrm{min}}~
||y(W, O) - \sum_{r,d} A(W, O; d,r) \xzero_{dr}||^2 + \alpha|| \xzero||^2,\label{eq:ls-mini}
\end{align}
which is a regularized convex optimization in the well-known form of
Ridge regression. The tensor $A$ reads
\begin{align}
& A(W, O; d,r)\equiv  B(W, r, d) x^{(3)}_{dO},
\end{align}
where
\begin{align}
B(W, r, d) &\equiv
\sum_{l_1}
U_{l_1}^{(r)}(i\omega)
x^{(1)}_{dl_1}
\sum_{l_2}
U_{l_2}^{(r)}(i\omega')
x^{(2)}_{dl_2}.
\end{align}
In matrix form, Eq.~(\ref{eq:ls-mini}) reads
\begin{align}
&\underset{\bx^{(0)}}{\mathrm{min}}~||\by - \bA \bx^{(0)}||^2 + \alpha ||\bx^{(0)}||^2,\label{eq:ls-mini-matrix-form}
\end{align}
where $\by$ and $\bx^{(3)}$ are flatted 1D arrays.
The matrix is size of $(N_W N_O, N_rD)$, where $N_r$ is the number of different representations (=12), $N_O$ is the size of the combined index for spin and orbitals.

\begin{figure}
  \begin{center}
    \resizebox{.75\columnwidth}{!}{%
      \begin{tikzpicture}[
        font=\sffamily,
        source/.style={draw,thick,rounded corners,fill=blue!20,inner sep=.3cm},
        ident/.style={draw,thick,inner sep=.3cm},
        process/.style={draw,thick,circle,inner sep=1pt, minimum size=22pt, fill=blue!20},
        every node/.style={align=center},
        bbprime/.style={draw,text width=2cm,minimum height=.75cm,minimum
          width=1.8cm,thick,
          rounded corners,fill=blue!20,inner sep=.3cm, anchor=center}
        ]
        \tikzstyle{myarrows}=[line width=1mm,-triangle 45, fill
        opacity=0., postaction={draw, line width=1mm, shorten >=1mm, -}]
        
        \node[bbprime] (full_correlator) {$B\, \mathrm{or}\,B'$};
        \node[process, fill=orange!60, below left=0.5 cm and -.6cm of full_correlator]
        (u1) {$u$};
        \node[process, below right=0.5 cm and +.22cm of full_correlator]
        (x3) {$x^{(3)}$};
        \node[ident, path picture={%
          \draw (path picture bounding box.north west) -- (path
          picture bounding box.south east);},
        below= 1.8cm of full_correlator] (i1) {};
        \node[circle,inner sep=0pt,minimum size=-0.1pt, above=.0cm of i1] (pointfusion) {};
        \draw[thick] (u1.90) -- node[pos=.4,right]
        {$r\, \mathrm{ or }\, l_1$} ++(0, .41);
        \draw[thick] (pointfusion) -- node[pos=.2,right] {$d$} ++(0, 1.8);
        \draw[thick] (pointfusion) -- (u1.-45);
        \draw[thick] (pointfusion) -- (x3.-135);
        \draw[thick] (full_correlator.90) -- node[pos=.7,left]
        {$\mathcal{W}$} ++(0,.6);
        \draw[thick] (x3.90) -- node[pos=.89,left]
        {$O$} ++(0,1.88);
        
        \node [right color=green!50!cyan,
        single arrow,
        minimum height=1.4cm,
        shading angle=90+60,
        inner sep=0.2cm,
        right=.9 cm of full_correlator] (arrowa) {};
        \node[below right=-.cm and -1.4 cm of arrowa]
        (u1dd) {$\scriptstyle \mathcal{O}(N_WDN_{r/l})$};
        
        \draw[thick, dashed]
        ($(full_correlator.north west)+(-0.5,0.15)$)
        rectangle ($(full_correlator.south east)+(0.05,-1.25)$);
        
        \node[above left=.25cm and . cm of full_correlator]
        () {(a)};

        \node[process, right=+2.4cm of full_correlator]
        (x3shadow) {};
        \node[process, right=+3.8cm of full_correlator]
        (x3prime) {$x^{(3)}$};
        \draw[thick] (x3shadow.90) -- node[pos=.7,left]
        {$\mathcal{W}$} ++(0,.6);
        \draw[thick] (x3prime.90) -- node[pos=.7,left]
        {$O$} ++(0,.6);
        \draw[thick] (x3shadow.0) -- node[pos=.5,above]
        {$d$} (x3prime.180);
        
        \node[bbprime, below right=1.8cm and -.8cm of x3shadow]
        (partial_correlator) {};
        \draw[thick] (partial_correlator.35) -- node[pos=.7,left]
        {$O$} ++(0,.6); 
        \draw[thick] (partial_correlator.145) -- node[pos=.7,left]
        {$\mathcal{W}$} ++(0,.6); 
        
        \node [right color=green!50!cyan,
        single arrow,
        minimum height=1cm,
        shading angle=90+60,
        rotate=-90,
        inner sep=0.2cm,
        below right=0.2cm and +3.7cm of full_correlator] (arrowb) {};
        \node[above right=0.cm and .1 cm of arrowb]
        (u1dd) {$\scriptstyle \mathcal{O}(N_W N_{O})$};

        \node[bbprime, below=5.5cm of full_correlator] (full_correlatorb) {$B\, \mathrm{or}\,B'$};
        \node[process, below right=0.5 cm and +.19cm of full_correlatorb]
        (x3b) {$x^{(3)}$};
        \node[ident, path picture={%
          \draw (path picture bounding box.north west) -- (path
          picture bounding box.south east);},
        below= 1.8cm of full_correlatorb] (i1b) {};
        \node[circle,inner sep=0pt,minimum size=-0.1pt, above=.0cm of i1b] (pointfusionb) {};
        \draw[thick] (pointfusionb) -- node[pos=.25,right]
        {$d$} ++(0, 1.8);
        \draw[thick] (pointfusionb) -- (x3b.-135);
        \draw[thick] (pointfusionb) -- ++(-.4,.4);
        \draw[thick] (full_correlatorb.90) -- node[pos=.7,left]
        {$\mathcal{W}$} ++(0,.6);
        \draw[thick] (x3b.90) -- node[pos=.89,left]
        {$O$} ++(0,1.88);
        
        \node[coordinate, below left=0.7 cm and -.4cm of full_correlatorb]
        (u1b) {};
        \draw[thick] (u1b.90) -- node[pos=.4,left]
        {$r\, \mathrm{ or }\, l_1$} ++(0, .71cm);

        \node[bbprime, minimum width=3.5cm, fill=orange!60,
        above right=.6cm and -2.6cm of full_correlatorb] (v) {$v$};

        \node [right color=green!50!cyan,
        single arrow,
        minimum height=1.2cm,
        shading angle=90+60,
        inner sep=0.2cm,
        right=1.1 cm of full_correlatorb] (arrowc) {};
        \node[above right=.cm and -1.1 cm of arrowc]
        (u1dd) {$\scriptstyle \mathcal{O}(N_W N_{O})$};
        
        \draw[thick, dashed]
        ($(v.north east)+(-.9,0.15)$)
        rectangle ($(x3b.south east)+(0.2, -.15)$);
        
        \node[above left=.25cm and . cm of v] () {(b)};
        
        \node[bbprime, right=2.4cm of full_correlatorb] (full_correlatorc) {$B\, \mathrm{or}\,B'$};
        \node[bbprime, minimum width=3.5cm, fill=blue!20,
        above right=.6cm and -2.6cm of full_correlatorc] (vshadow) {};
        \node[ident, path picture={%
          \draw (path picture bounding box.north west) -- (path
          picture bounding box.south east);},
        below right= .5cm and .3cm of full_correlatorc] (i1c) {};
        \node[circle,inner sep=0pt,minimum size=-0.1pt, above=.0cm of i1c] (pointfusionc) {};
        \draw[thick] (pointfusionc) -- node[pos=.75,right]
        {$d$} ++(0, 2.);
        \draw[thick] (pointfusionc) -- (full_correlatorc.-25);
        \draw[thick] (pointfusionc) -- ++(.4,.4);
        \draw[thick] (full_correlatorc.90) -- node[pos=.7,left]
        {$\mathcal{W}$} ++(0,.6);
        \node[coordinate, below left=0.7 cm and -.4cm of full_correlatorc]
        (u1c) {};
        \draw[thick] (u1c.90) -- node[pos=.4,left]
        {$r\, \mathrm{ or }\, l_1$} ++(0, .71cm);
        
        \node[bbprime, below=2.4cm of full_correlatorc] (full_correlatord) {};
        \node[coordinate, below left=0.7 cm and -.4cm of full_correlatord]
        (u1d) {};
        \draw[thick] (u1d.90) -- node[pos=.4,left]
        {$r\, \mathrm{ or }\, l_1$} ++(0, .71cm);
        \node[coordinate, below right=0.7 cm and -.4cm of full_correlatord]
        (u1dd) {};
        \draw[thick] (u1dd.90) -- node[pos=.4,left]
        {$d$} ++(0, .71cm);
        
        \node [right color=green!50!cyan,
        single arrow,
        minimum height=1.7cm,
        shading angle=90+60,
        rotate=-90,
        inner sep=0.2cm,
        below right=1.cm and +4.cm of full_correlatorb] (arrowd)
        {};
        \node[above right=-.3cm and .1 cm of arrowd]
        (u1dd) {$\scriptstyle \mathcal{O}(N_W N_{r/l})$};
        
      \end{tikzpicture}
    }%
  \end{center}
  \caption{Graphical representations of tensor contractions for
    solving Eqs.~(\ref{eq:ls-mini}) and (\ref{eq:ls-mini-xone}),
    iteratively.
    The panel (a) illustrates the order of tensor contractions for computing $\bA \bu$ and $\bA^\prime \bu$.
    The panel (b) illustrates the procedure for computing $\bA^\dagger \bv$ and $(\bA^\prime)^\dagger \bv$.
    The computational complexity for each operation is shown.
  }
  \label{fig:ALS1}
\end{figure}
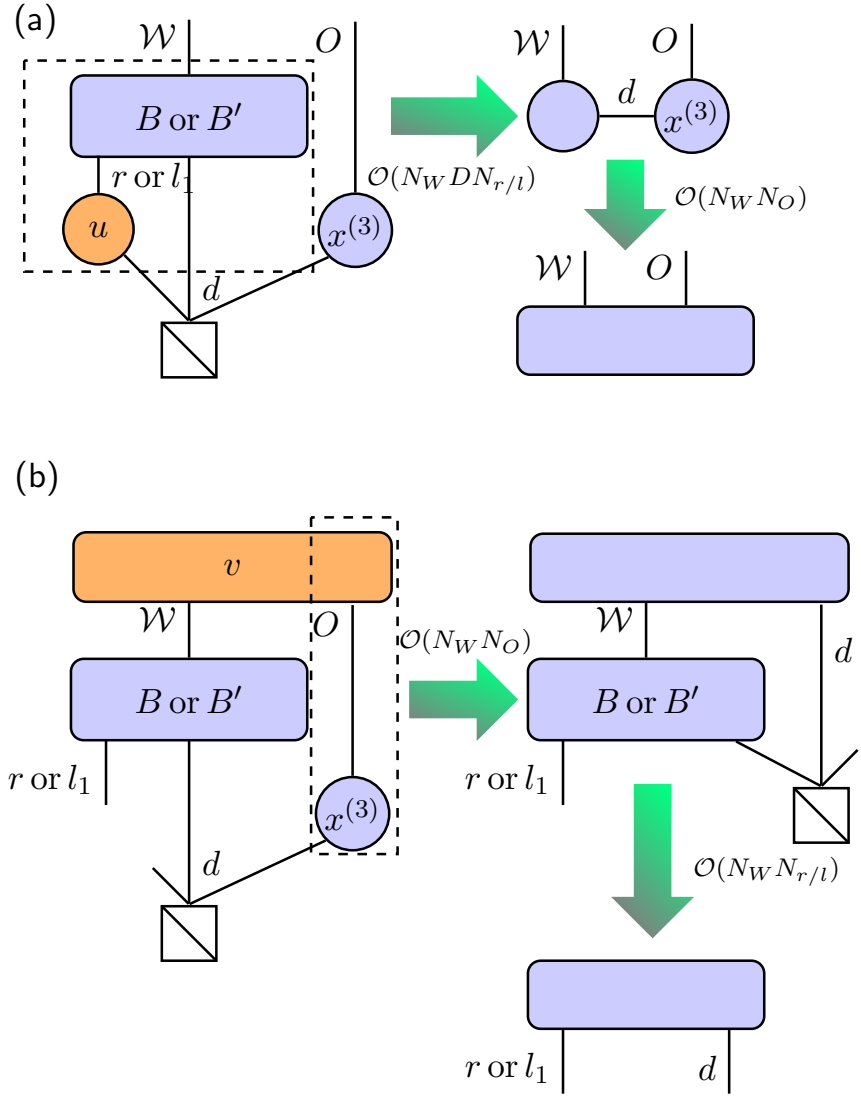

In practice, we solve Eq.~(\ref{eq:ls-mini-matrix-form}) by an iterative method, LSQR~\cite{LSQR}, without constructing the matrix $\bA$ explicitly.
In LSQR, the matrix $\bA$ is used only to compute $\bA \bv$ and $\bA^\dagger \bu$ for various $\bv$ and $\bu$.
Hence we store the (precomputed) tensor $B$ in memory, and compute these products by means of tensor contractions.
We illustrate the tensor contractions for computing $\bA \bv$ and $\bA^\dagger \bu$ in Figs.~\ref{fig:ALS1}(a) and~\ref{fig:ALS1}(b), respectively.
This approach not only reduces memory footprints but also reduces the computational complexity from $O(N_W D N_r N_O)$ to $O(N_W D N_r) + O(N_W D N_O)$.

\subsubsection{Optimization of $\xone$}
The optimization of $\xone$ can be done in a way very similar to that of $\xzero$.
Thus, we focus only on the differences.
The reduced least squares problem reads
\begin{align}
\underset{\xone}{\mathrm{min}}~
||y(W, O) - \sum_{d, l_1} A^\prime(W, O; d, l_1) \xone_{d l_1}||^2 + \alpha|| \xone||^2,\label{eq:ls-mini-xone}
\end{align}
where
\begin{align}
& A^\prime(W, O; d, l_1)\equiv  B^\prime(W, d, l_1) x^{(1)}_{dO}
\end{align}
where
\begin{align}
B^\prime(W, d, l_1) &\equiv
x_{dr}^{(0)}
\sum_{l_1}
U_{l_1}^{(r)}(i\omega)
x^{(1)}_{dl_1}.
\end{align}
As illustrated in Fig.~\ref{fig:ALS1}, we can readily compute $\bA^\prime \bv$ and ${\bA^\prime}^\dagger \bu$ by tensor contractions.

\subsubsection{Optimization of $\xtwo$}
The tensor $\xtwo$ can be optimized in exactly the same way as $\xone$. Thus, we do not describe the optimization of $\xtwo$ for simplicity.

\subsubsection{Optimization of $\xthree$}
The optimization of $\xthree$ is rather simple.
The reduced least squares problem reads
\begin{align}
\underset{\xthree}{\mathrm{min}}~
||y(W, O) - \sum_{d,O} A^{\prime\prime}(W; d, O) \xthree_{d O}||^2 + \alpha|| \xthree||^2,\label{eq:ls-mini-xthree}
\end{align}
where the tensor $A^{\prime\prime}$ can be stored in memory.
Furthermore, this optimization problem is separable with respect to $D$ and thus can be solved independently.

\subsection{Acceleration techniques and convergence condition}
In the previous subsection, we have explained how to perform one sweep through the tensors.
This single ALS sweep corresponds to the function ALS in Algorithm~\ref{algorithm:ALS}.
The function ALS takes an array obtained by flattening tensors of fitting parameters as input.
After a single sweep, the updated tensors are returned as a flattened array.
Here, flattening means recasting multiple tensors of complex numbers
into a single one-dimensional array of real numbers (the order is
arbitrary).

The whole procedure of the accelerated ALS is illustrated in Algorithm~\ref{algorithm:ALS}.
The main difference from the plain ALS is the existence of $\beta$.
In the second last line,
$\beta (\boldsymbol{x}_k - \boldsymbol{x}_{k-1})$ acts as a momentum term for
$\beta>0$.
Although this momentum term accelerates the convergence by updating
the parameters aggressively, this sometimes leads to oscillatory
behavior or divergence.
We use a restarting mechanism to stabilize the accelerated ALS.
In practice,
when the restarting condition $f(\boldsymbol{x}_{k}) > f(\boldsymbol{x}_{k-1})$ is met ($f$ is the cost function),
a ALS sweep is forced by setting $\beta=0$ (see the comment in Algorithm~\ref{algorithm:ALS}).
For more details on the acceleration techniques, please refer to Ref.~\cite{mitchell2018}.

The loop is exited when a convergence condition is met.
$\delta$ in Algorithm~\ref{algorithm:ALS}) is a relative tolerance.

\subsection{Technical details and numerical results}
We parallelize the whole fitting procedure by MPI with respect to frequencies.
This parallelization is efficient particularly for $\Gfour$.
We parallelize the LSQR implementation in the SciPy Python package\cite{scipy01} using MPI with respect to sampling frequencies.

\begin{figure}
	\centering
	\includegraphics[width=0.7\linewidth]{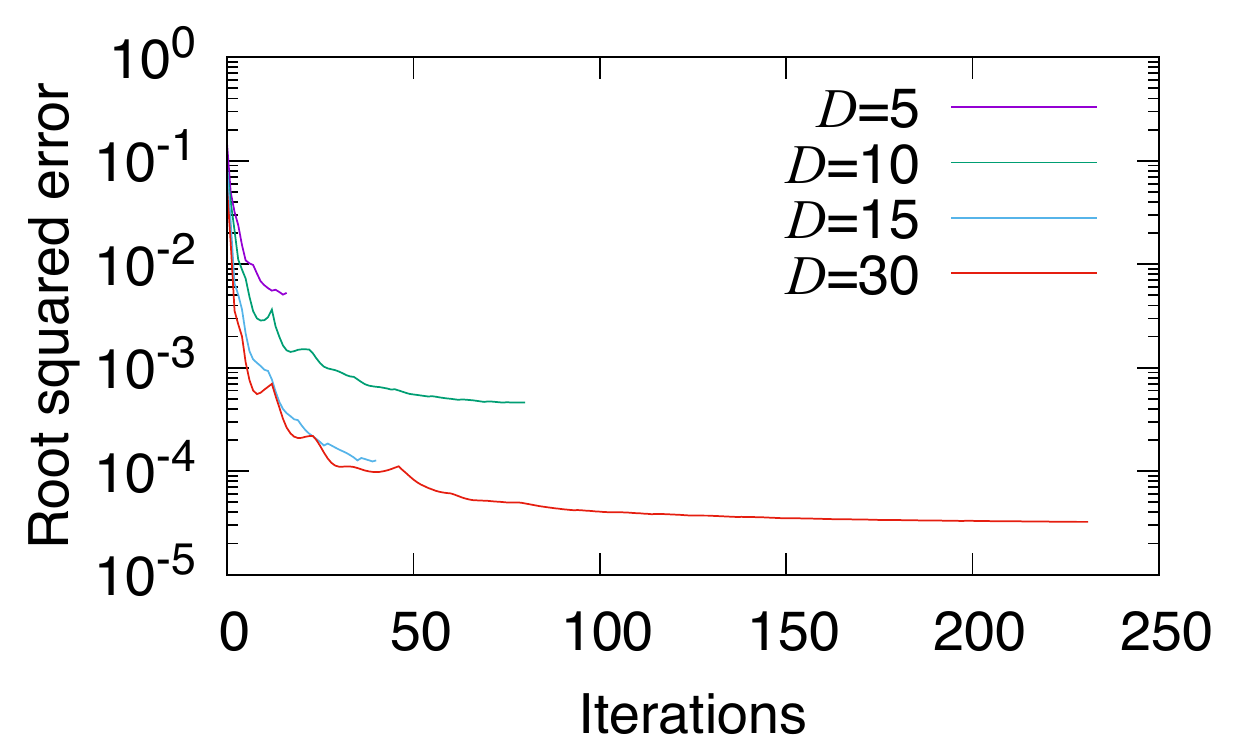}
	\caption{Convergence of root squared errors for the local
		susceptibility $X^\mathrm{loc}$ of the single-band Hubbard model
		analyzed in Section~\ref{sec:single-band-hubbard-model}.
	}
	\label{fig:convergence_fittig}
\end{figure}

Figure~\ref{fig:convergence_fittig} shows the convergence of the root
squared errors of the local susceptibility for the single-band Hubbard
model analyzed in Sec.~\ref{sec:single-band-hubbard-model}.
One can see that the fitting errors quickly converge.
The small oscillatory behavior is due to the acceleration.

\begin{algorithm}[H]
	\begin{algorithmic}
		\Function{ALS}{$\boldsymbol{x}$}
		\State $\{x^{(i)}\} \gets \textrm{UNFLATTEN}(\boldsymbol{x})$ \Comment{Unflatten input array}
		\For{$i = 0, 1, 2, 3$}
		\State Construct $A^{(i)}$
		\State $x^{(i)} \gets \mathrm{LSQR}(y, A^{(i)}, \alpha)$
		\EndFor
		\State Return \textrm{FLATTEN}($\{x^{(i)}\}$) \Comment{Flatten updated tensors}
		\EndFunction
		
		\State
		
		\State $\boldsymbol{x}_0 \gets \text{Random~initial values}$ 
		\State $\boldsymbol{x}_{1} \gets \mathrm{ALS}(\boldsymbol{x}_0)$
		\For{$k = 1, 2, ...$}
		
		\If{$f(\boldsymbol{x}_{k}) > f(\boldsymbol{x}_{k-1})$}
		\State $\boldsymbol{x}_{k} \gets \boldsymbol{x}_{k-1}$
		\State $\beta \gets 0$ \Comment{Force ALS at this iteration}
		\Else
		\State $\beta \gets 1$
		\EndIf
		\State $\boldsymbol{x}_{k+1} \gets \mathrm{ALS}(\boldsymbol{x}_k + \beta (\boldsymbol{x}_k - \boldsymbol{x}_{k-1})   )$
		\If{$|f(\boldsymbol{x}_{k+1})-f(\boldsymbol{x}_{k})| < \delta \times f(\boldsymbol{x}_{k})$}
		\State Exit loop
		\EndIf
		\EndFor
	\end{algorithmic}
	\caption{Accelerated alternating least squares}
	\label{algorithm:ALS}
\end{algorithm}

\end{appendix}


\nolinenumbers

\end{document}